\documentclass[12pt,fleqn]{article}

\usepackage{amsmath,amsfonts,bm,amssymb}
\usepackage{amsthm}

\usepackage{ifpdf}

\usepackage{eucal}
\usepackage{eufrak}
\usepackage[margin=2cm,a4paper]{geometry}
\usepackage{url}
\usepackage{multirow}
\usepackage{rotating}
\usepackage{graphicx}
\usepackage{grffile}
\usepackage{color}
\usepackage{subfig}

\newcommand{\abs}[1]{\left\lvert #1 \right\rvert}
\newcommand{\ud}{\mathrm{d}}
\newcommand{\ue}{\mathrm{e}}

\long\def\symbolfootnote[#1]#2{\begingroup\def\thefootnote{\fnsymbol{footnote}}\footnote[#1]{#2}\endgroup}

\newcommand{\captionfonts}{\footnotesize}

\makeatletter
\long\def\@makecaption#1#2{%
  \vskip\abovecaptionskip
  \sbox\@tempboxa{{\captionfonts #1: #2}}%
  \ifdim \wd\@tempboxa >\hsize
    {\captionfonts #1: #2\par}
  \else
    \hbox to\hsize{\hfil\box\@tempboxa\hfil}%
  \fi
  \vskip\belowcaptionskip}
\makeatother

\begin{document}
 
\title{Multiplicative noise, fast convolution, and pricing}

\author{Giacomo Bormetti\,$^{\textrm{a, b}}$, Sofia Cazzaniga\,$^{\textrm{c}}$} 

\date{\today}
\maketitle
\small
\begin{center}
  $^\textrm{a}$~\emph{Scuola Normale Superiore, Piazza dei Cavalieri 7 Pisa I-56126, Italy}\\
  $^\textrm{b}$~\emph{INFN, Sezione di Pavia, via Bassi 6 Pavia I-27100, Italy}\\
  $^\textrm{c}$~\emph{Swiss Finance Institute at the University of Lugano, Via Buffi 13 Lugano, Switzerland}
\end{center}
\normalsize

\vspace{0.3cm}

\begin{abstract}
	In this work we detail the application of a fast convolution algorithm computing high dimensional integrals to the context of
	multiplicative noise stochastic processes. The algorithm provides a numerical solution to the problem of characterizing
	conditional probability density functions at arbitrary time, and we applied it successfully to quadratic and piecewise linear 
	diffusion processes. The ability in reproducing statistical features of financial return time series, such as thickness of the
	tails and scaling properties, makes this processes appealing for option pricing. Since exact analytical results are missing, 
	we exploit the fast convolution as a numerical method alternative to the Monte Carlo simulation both in objective and risk neutral 
	settings. In numerical sections we document how fast convolution outperforms Monte Carlo both in velocity and efficiency terms. 
\end{abstract}

\smallskip

\smallskip


\section{Introduction}\label{sec:intro}
Two of the basic problems computational finance has to deal with are the choice of the optimal model driving the stochastic 
evolution of financial variables and, once a good candidate has been identified, the search of a reliable way for its fast and accurate 
simulation. The former issue has been widely investigated both by econometricians, mathematicians, and physicists, as demonstrated by the increasing 
literature on this topic, see for example~\cite{Campbell_etal, Mandelbrot, Mantegna_Stanley, Bouchaud_Potters, McCauley}.
Tracing back to the work of~\cite{Mandelbrot.1963} and the analysis in~\cite{Fama}, 
empirical studies have shown that financial time series exhibit features departing from the Gaussian assumption.
In~\cite{Cont} a detailed review of the stylized empirical facts emerging in various type of financial markets
is presented and discussed. These findings are nowadays accepted as universal evidences,
shared among different markets in different epochs. From the earlier results on cotton prices of Mandelbrot or the thick tailed nature
of the Dow Jones Industrial Average recognized by Fama, very heterogeneous models have been proposed in order 
to reproduce the degree of asymmetry and the excess of kurtosis of the empirical distributions.
Approaches directly developing from distributional assumptions includes the truncated L\'evy model discussed in~\cite{Mantegna_Stanley}, and 
those employing generalized Student-$t$ and exponential distributions, see~\cite{Bouchaud_Potters, McCauley_Gunaratne}.
Different mechanisms also capturing the observed non trivial structure of higher order correlation functions
model the stochastic nature of the return volatility. 
Continuous time approaches have been extensively analyzed and range from the fractional Brownian motion,~\cite{Mandelbrot},
to stochastic volatility models, for a review we suggest~\cite{Fouque_Papanicolaou_Sircar}.  
Discrete time models include AutoRegressive Conditional Heteroskedastic (ARCH) and Generalized ARCH processes,~\cite{Engle,Bollerslev}, 
and multifractal ones,~\cite{Borland_etal}, the latter being inspired by cascades originally introduced by Kolmogorov in the context of turbulent flows. 
Turbulent velocity flows have also led to a series of empirical works testing and strongly relying on the Markovian 
nature of foreign exchange returns,~\cite{Ghashgaie_etal, Friedrich_etal}. The macroscopic description of the observed phenomena is
provided in terms of a Fokker-Planck~(FP) equation with linear drift and quadratic diffusion coefficients. Processes leading to an
equation with the same structure characterize several physical systems, as reviewed in~\cite{Bormetti_Delpini}. 
Also the statistical feedback mechanism proposed in~\cite{Borland-PRL, Borland-QF} can be recast
in terms of non linear diffusion, as originally remarked by~\cite{McCauley_etal.2007a}, who also pointed out potential problems arising 
when computing expectations under power law tailed distributions. Even though these drawbacks have been later amended in~\cite{Vellekoop-Nieuwenhuis}, 
\cite{McCauley_etal.2007b} propose to switch to exponential tailed PDFs. In particular, they develop a general approach to generate a Markovian process 
obeying scaling relations starting from a driftless stochastic differential equation (SDE). In the current paper we focus on
the numerical characterization of these latter processes, of the above mentioned quadratic diffusion ones, and in their application in financial modeling.
To this respect, especially for pricing purposes, we need to reconstruct the conditional probability
density function (PDF) describing the stochastic dynamics in order to evaluate expectations of future payoffs. 
Yet a closed form expression for the density is rarely available. For this reason several numerical procedures have been developed and 
have become the common practice, e.g. binomial and multinomial lattice algorithms, Monte Carlo (MC) simulation, and partial differential equation solvers
(for a review see~\cite{Brandimarte}). We decide to investigate and widely exploit the fast convolution algorithm (FCA) introduced in~\cite{Eydeland}. 
The algorithm applies to Markovian stochastic processes: the repeated application of the Chapman-Kolmogorov equation, and a clever problem re-formulation
allow to rewrite functional integrals in terms of Fourier and anti-Fourier transforms of the state vector. 
Performing these operations via fast Fourier transform (FFT) algorithm, numerical efficiency is achieved and computational complexity is notably reduced.

The structure of the paper is the following. After introducing stochastic models we have chosen to investigate, we detail step by step FCA; 
in paragraph~\ref{subsec:nr-mnoise} we test its numerical performances against the standard MC approach for different specification of models and
parameter values. Section~\ref{sec:fin_application} is dedicated to financial applications in the context of option pricing. 
In~\ref{subsec:riskneutral} we derive the risk neutral measure for the piecewise diffusion process of~\cite{McCauley_Gunaratne}, and 
the exact formula for Plain Vanilla pricing. We then consider geometric Asian options by thoroughly develop the two dimensional setting 
required by fast convolution, see paragraph~\ref{subsec:asian}. We exploit the formal analogy between the latter case and the framework 
discussed by~\cite{Vellekoop-Nieuwenhuis, Borland-PRL, Borland-QF} to price Plain Vanilla options in 
paragraph~\ref{subsec:VNB_model}, and in the final part we collect numerical distributions and implied volatility surfaces we have obtained
to prove the reliability of FCA. We draw relevant conclusions and possible perspectives in section~\ref{sec:conclusions}.

\section{Stochastic processes with multiplicative noise}\label{sec:mnoise}
Multiplicative stochastic processes we investigate in this work correspond to the class of 
quadratic diffusion described in~\cite{Bormetti_Delpini, Delpini_Bormetti} and of piecewise linear diffusion 
introduced in~\cite{McCauley_Gunaratne} and later on rediscussed by~\cite{Alejandro_etal} in a slightly different flavour. 

The SDE describing the quadratic diffusion dynamics under It\^o prescription can be written as
\begin{equation}\label{eq:quadraticSDE}
	\ud X_t=\frac{aX_t+b}{g(t)}\ud t + \sqrt{\frac{cX_t^2+dX_t+e(t)}{g(t)}}\ud W_t\, ,
\end{equation}
with $X_{t_0}=0$ initial time condition; $\ud W_t$ is the standard Brownian increment, $1/g(t)$ and $e(t)$ are non-negative smooth functions for $t\geq t_0$.
We require $\mathrm{D}^2\doteq 4~c~e(t)-d^2\geq 0$ with $c>0$.
Equation~\eqref{eq:quadraticSDE} governs the dynamics of a variety of complex phenomena as reviewed in~\cite{Bormetti_Delpini}, where
an exhaustive description of the process in terms of its moments is also provided. For instance, when $e(t)=e$ is constant, 
it is possible to characterize analytically the time-scaling adjustment of the process $X_t$, regulated by the choice of $g(t)$, and its convergence to the stationary state. 
If $a$ is non negative or if $e$ is time dependent, $X_t$ lacks stationarity; however, in the final section of this paper we will discuss an 
application of this latter case to the context of financial option pricing, tracing back to~\cite{Borland-PRL, Borland-QF}, and later revised by~\cite{Vellekoop-Nieuwenhuis}.

The quadratic diffusion process~(\ref{eq:quadraticSDE}) can be formally manipulated to reduce it in a more convenient form 
by means of the Lamperti transform, as we will see in Section~\ref{subsec:FCA}. Here, we slightly simplify it introducing a new variable beating the time 
$\tau(t)=\int_{t_0}^{t}\ud s/g(s)$, which we will refer to as  the integral time. In this new setting the process $X_\tau$ is described by the following dynamics
\begin{equation}\label{eq:quadraticSDEtau}
	\ud X_\tau = (aX_\tau+b)\ud\tau+\sqrt{cX_\tau^2+dX_\tau+\tilde{e}(\tau)}\ud W_\tau\, ,
\end{equation}
with $X_0=0$ and $\tilde{e}(\tau)=e(t(\tau))$\footnote{By virtue of the properties of $g$, $\tau$ is a monotonously increasing function of $t$, 
implying the well-definiteness of the inverse function $t(\tau)$.}.

As far as piecewise linear diffusion is concerned, the main property we are interested in is scaling, relating returns over different sampling intervals. 
More precisely, whenever returns are rescaled by a factor $t^H$, the shape of their distribution scales according to 
\begin{equation}\label{eq:scaling}
	P(x, t)=\frac{1}{t^H}~G\left(\frac{x}{t^H}\right),
\end{equation}
where $G$ is the so called scaling function.
Following~\cite{Alejandro_etal}, piecewise diffusion process can be defined by means of the driftless SDE
\begin{equation}\label{eq:piecewiseSDE}
	\ud X_t = \sigma\sqrt{1+\epsilon\frac{\abs{X_t}}{t^H}}\ud W_t\, ,\quad\mathrm{with}\quad X_0=0\,.
\end{equation}
Through the analysis of the FP equation satisfied by $P(x,t)$, it can be readily shown that $1/2$ is the only value of $H$ consistent with the scaling 
assumption~\eqref{eq:scaling}. In addition to this the FP equation admits the following solution
\begin{equation}\label{eq:exponential}
	P(x, t)=\frac{\ue^{-\alpha}}{2\sigma^{2\alpha}\epsilon^\alpha\Gamma[\alpha, \alpha ]\sqrt{t}}\exp\left[-\frac{|x|}{\sigma^2\epsilon\sqrt{t}}\right]
	\left(1+\epsilon\frac{|x|}{\sqrt{t}}\right)^{\alpha-1}\, ,
\end{equation}
where $\alpha=1/(\sigma^2\epsilon^2)$ and $\Gamma[a, z]=\int_{z}^{\infty}s^{a-1}e^{-s}\ud s$.
The interest in previous density is twofold, both because of the emergence of a scaling exponent $1/2$,
in agreement with the empirical findings, and because of the evidence of leptokurtosis.
~\cite{McCauley_Gunaratne, McCauley} derive closed-form option pricing formulae under the distribution~\eqref{eq:exponential}; 
yet they show that consistency between scaling, exponential PDF and martingale option pricing requires the replacement of the 
It\^o correction for $X_t$ under the risk neutral measure with a constant, see section 2 in~\cite{McCauley_etal.2007b}. 
We argue this approximation is questionable, and we want to perform option pricing not relying on it, therefore we need a numerical methodology
whose efficiency and flexibility promise to compensate for the absence of closed-form solution. 

\subsection{Fast convolution algorithm}\label{subsec:FCA}
In this section we review the fast convolution algorithm proposed in \cite{Eydeland}.

Let us consider the generic process $X_\tau$, whose dynamics is described by the following general SDE
\[
\ud X_\tau = \mathrm{M}_X(X_\tau, \tau) \ud \tau + \mathrm{D}_X (X_\tau,\tau) \ud W_\tau\, ,\quad X_{\tau=0}=X_0\,.
\]
We start by transforming the process $X_\tau$ into one with unitary diffusion coefficient. This is performed via the
Lamperti transform, see section 1.11.4 in~\cite{Iacus}, defined as
\[
Z_\tau(X_\tau,\tau)=\int_{X_0}^{X_\tau}\frac{\ud\hat{X}}{\mathrm{D}_X(\hat{X},\tau)}\,.
\]
Under suitable regularity condition It\^{o} Lemma can be applied to $Z_\tau(X_\tau,\tau)$ and its dynamics turns out to be
\begin{equation}\label{eq1}
	\ud Z_\tau=\mathrm{M}_Z(Z_\tau,\tau) \ud \tau + \ud W_\tau\, ,\quad Z_{\tau=0}=0\, ,
\end{equation}
with 
\[
\mathrm{M}_Z(Z_\tau,\tau)=\frac{\tilde{\mathrm{M}}_X(X(Z_\tau),\tau)}{\tilde{\mathrm{D}}_X(X(Z_\tau),\tau)}
+\frac{\partial}{\partial \tau}\int_{X_0}^{X(Z_\tau)}\frac{\ud\hat{X}}{\mathrm{D}_X(\hat{X},\tau)}-\frac{1}{2}
\tilde{\mathrm{D}}_X(X(Z_\tau),\tau)\, ,
\]
where $\tilde{\mathrm{M}}_X$ is the function $\mathrm{M}_X(X_\tau,\tau)$ evaluated in $X(Z_\tau)$, and analogously for $\tilde{\mathrm{D}}_X$.
Our aim is to provide an approximate expression for the transition probability density function $p(z_\tau,\tau|z_0,0)$\footnote{From now on 
we will drop the explicit dependence on the time variable $\tau$.}.
We introduce an equally spaced time grid $0=\tau^0, \tau^1, \dots, \tau^n=\tau$, with $\tau^i=i\Delta \tau$, 
in a similar spirit to the path integral approach, see \cite{Dash,Montagna-etal-PHYA,Bormetti-etal-QF,Baaqui}. 
The repeated use of the Chapman-Kolmogorov equation in this discrete setting allows to write the transition probability for a generic $\tau>0$
as a finite high dimensional integral 
\begin{eqnarray}\label{eq:iterCK}
	p(z^n|z^0)\simeq \int_{z^n}\int_{z^{n-1}}\dots\int_{z^1} \prod_{i=1}^{n-1}\ud z^i~\pi(z^n|z^{n-1})\pi(z^{n-1}|z^{n-2})\dots\pi(z^1|z^0)\, ,
\end{eqnarray}
where $z^i=z(\tau^i)$, and $\pi$ is the short time transition PDF that we chose equal to the Normal density 
\[
\pi(z^{i+1}|z^{i})=\frac{1}{\sqrt{2\pi\Delta\tau}}\exp\left[-\frac{(z^{i+1}-z^i-\mathrm{M}_Z(z^i,\tau^i)\Delta\tau)^2}{2\Delta\tau}\right]\, .
\]
By means of the new variables $\xi^i\doteq z^i+\mathrm{M}_Z(z^i,\tau^i)\Delta\tau$, the transition becomes 
symmetric under the exchange of $z^{i+1}$ with $\xi^i$, i.e. $\tilde\pi(z^{i+1}|z^i(\xi^i))=\tilde\pi\left( (z^{i+1}-\xi^i)^2\right)$.
For each one dimensional integration appearing in equation~(\ref{eq:iterCK}), we have
\begin{equation}\label{eq:1Dinteg}
	p(z^{i+1}|z^0) = \int_{z^i} \ud z^i~\pi(z^{i+1}|z^{i})p(z^i|z^0) = \int_{\xi^i} \ud \xi^i 
	\frac{\partial z^i}{\partial\xi^i}\tilde\pi\left((z^{i+1}-\xi^{i})^2\right)\tilde p(z^i(\xi^i)|z^0)\, ,
\end{equation}
where $\tilde p(z^i(\xi^i)|z^0)$ is the density $p(z^i|z^0)$ evaluated in $z^i(\xi^i)$, and similarly for $\tilde\pi$.
If we introduce a numerical integration grid of equally spaced points $z^{i}_j = \xi^i_j = z_\mathrm{min} + j\Delta z$ for all $i=0,\dots, n$ and $j=0,\dots, m-1$, 
where neither $z_\mathrm{min}$ nor $\Delta z$ depend on the time label $i$, then 
the PDF $\tilde\pi(z^{i+1}_{j'}|\xi^i_{j})$ associated to transition of moving from point $\xi_{j}$ at time $\tau^i$ to point $z_{j'}$ at time
$\tau^{i+1}$ is function only of the difference $j'-j$, i.e. $\tilde\pi_{j'j}\doteq\tilde\pi\left( (j'-j)^2\Delta z^2 \right)$.  
The discrete matrix of transition probabilities 
\begin{eqnarray}
	\tilde{\Pi}=\left[ 
	\begin{array}{cccc} 
		\tilde\pi_{0 0}&\tilde\pi_{0 1}&\ldots&\tilde\pi_{0 (m-1)}\\
		\tilde\pi_{1 0}&\tilde\pi_{1 1}&\ldots&\tilde\pi_{1 (m-2)}\\
		\vdots&\vdots&\ddots&\vdots\\
		\tilde\pi_{(m-1) 0}&\tilde\pi_{(m-1) 1}&\ldots&\tilde\pi_{(m-1) (m-1)}
	\end{array}\right]\, ,
\end{eqnarray}
is therefore a symmetric Toeplitz matrix $\tilde\Pi_{ij}=\tilde\Pi_{|i-j|}$, with no dependence on the time variable. 
Letting
\begin{eqnarray}
	\boldsymbol{\mathrm{P}}^{i+1}=\left[ 
	\begin{array}{c} 
		p(z_0^{i+1}|z^0)\\
		p(z_1^{i+1}|z^0)\\
		\vdots\\
		p(z_{m-1}^{i+1}|z^0)
	\end{array}
	\right]\, ,\,
	\mathrm{J}^{i}=\left[ 
	\begin{array}{cccc} 
		\frac{\partial z_0^i}{\partial\xi_0^i}&0&\ldots&0\\
		0&\frac{\partial z_1^i}{\partial\xi_1^i}&\ldots&0\\
		\vdots&\vdots&\ddots&\vdots\\
		0&0&\ldots&\frac{\partial z_{m-1}^i}{\partial\xi_{m-1}^i}
	\end{array}\right]\, ,\,
	\tilde{\boldsymbol{\mathrm{P}}}^{i}=\left[ 
	\begin{array}{c} 
		\tilde{p}(z_0^{i}(\xi_0^i)|z^0)\\
		\tilde{p}(z_1^{i}(\xi_1^i)|z^0)\\
		\vdots\\
		\tilde{p}(z_{m-1}^{i}(\xi_{m-1}^i)|z^0)
	\end{array}
	\right]\, ,
\end{eqnarray}
equation~\eqref{eq:1Dinteg} can be approximated as
\[
\mathrm{P}_j^{i+1} \simeq \Delta z \sum_{k,l=0}^{m-1} \tilde\Pi_{jk} \mathrm{J}_{kl}^{i}\tilde{\mathrm{P}}_l^{i}\, . 	
\]
The entries of $\tilde{\boldsymbol{\mathrm{P}}}^{i}$ are computed by means of the linear interpolation operator 
\begin{eqnarray}
	\mathrm{I}^{i}=\frac{1}{\Delta z}\left[ 
	\begin{array}{ccccccccccc} 
		z_1^i-\xi_0^i&&\xi_0^i-z_0^i&&0&&\ldots&&0&&0\\
		0&&z_2^i-\xi_1^i&&\xi_1^i-z_1^i&&\ldots&&0&&0\\
		\vdots&&\vdots&&\vdots&&\ddots&&\vdots&&\vdots\\
		0&&0&&0&&\ldots&&z_{m-1}^i-\xi_{m-2}^i&&\xi_{m-2}^i-z_{m-2}^i\\
		0&&0&&0&&\ldots&&z_{m-1}^i-\xi_{m-1}^i&&\xi_{m-1}^i-z_{m-2}^i
	\end{array}\right]\end{eqnarray} 
	applied to $\boldsymbol{\mathrm{P}}^i$. 
Similarly, equation~\eqref{eq:iterCK} becomes
\begin{equation}\label{eq:numerical_CK}
	\boldsymbol{\mathrm{P}}^{n}\simeq
	\left(\Delta z\right)^{n-1}\tilde{\Pi}\mathrm{J}^{n-1}\mathrm{I}^{n-1}\ldots\tilde{\Pi}
	\mathrm{J}^{2}\mathrm{I}^{2}\tilde{\Pi}\mathrm{J}^{1}\mathrm{I}^1\boldsymbol{\mathrm{P}}^{1}\, .
\end{equation}
Matrix multiplications in previous equation are extremely time consuming. Indeed, while multiplying a $m$-vector by the $m\times m$ diagonal matrix $\mathrm{J}$ 
requires $m$ operations, and analogously for the $\mathrm{I}$ operator, multiplication of the $\tilde{\Pi}$ matrix by a vector requires $m^2$ operations. 
On top of this, the procedure must be repeated at each time step. As a consequence, the dominant contribution grows as $n\times m^2$, 
and choosing a rather thick grid, computational times rapidly explodes.
However, the multiplication of a Toeplitz matrix by a vector can be efficiently performed exploiting algorithms coming from digital signal processing.
By embedding $\tilde{\Pi}$ into a circulant matrix of dimensions $2m\times 2m$
\begin{eqnarray}
	\mathrm{C}=\left[ 
	\begin{array}{ccccccccc} 
		\tilde{\pi}_0&\tilde{\pi}_1&\dots&\tilde{\pi}_{m-1}&0&\tilde{\pi}_{m-1}&\tilde{\pi}_{m-2}&\dots&\tilde{\pi}_1\\
		\tilde{\pi}_1&\tilde{\pi}_0&\dots&\tilde{\pi}_{m-2}&\tilde{\pi}_{m-1}&0&\tilde{\pi}_{m-1}&\dots&\tilde{\pi}_2\\
		\vdots&\vdots&\ddots&\vdots&\vdots&\vdots&\ddots&\ddots&\vdots\\
		\tilde{\pi}_{m-1}&\tilde{\pi}_{m-2}&\dots&\tilde{\pi}_0&\tilde{\pi}_1&\tilde{\pi}_{2}&\dots&\dots&0\\
		0&\tilde{\pi}_{m-1}&\dots&\tilde{\pi}_1&\tilde{\pi}_0&\tilde{\pi}_1&\dots&\dots&\tilde{\pi}_{m-1}\\
		\tilde{\pi}_{m-1}&0&\dots&\tilde{\pi}_2&\tilde{\pi}_1&\tilde{\pi}_0&\tilde{\pi}_1&\dots&\tilde{\pi}_{m-2}\\
		\vdots&\vdots&\ddots&\vdots&\vdots&\vdots&\ddots&\ddots&\vdots\\
		\tilde{\pi}_1&\tilde{\pi}_2&\dots&0&\tilde{\pi}_{m-1}&\tilde{\pi}_{m-2}&\dots&\tilde{\pi}_1&\tilde{\pi}_0
	\end{array}
	\right]\, ,
\end{eqnarray}
the product of $\tilde{\Pi}$ with a generic vector $\boldsymbol{\mathrm{v}}$ is equal to the first $m$ components of 
$\mathrm{C}\boldsymbol{\mathrm{v}}_\ue$, $\boldsymbol{\mathrm{v}}_\ue\in\mathbb{R}^{2m}$, 
$\boldsymbol{\mathrm{v}}_\ue\doteq\left(\boldsymbol{\mathrm{v}}^\mathrm{t}, 0,\ldots, 0\right)^\mathrm{t}$.
Every circulant matrix can be expressed as $\mathrm{C}=\mathrm{U}\Lambda\mathrm{U}^*$, where $\mathrm{U}^*$ denotes conjugate transpose 
of $\mathrm{U}$, whose columns are $\mathrm{U}^{j}=(1,\ue^{-\pi\mathrm{i}j/m},\ldots,\ue^{-\pi\mathrm{i}j(2m-1)/m})^\mathrm{t}/\sqrt{2m}$ 
for $j=0,\ldots,2m-1$, and $\Lambda=\mathrm{diag}(\boldsymbol{\mathrm{C}}_{0})$, with $\boldsymbol{\mathrm{C}}_{0}$ first row of the circulant matrix.
Thanks to this result, the product $\mathrm{C}\boldsymbol{\mathrm{v}}_\ue$ can be performed exploiting fast Fourier transform (FFT) algorithm
\[
\mathrm{C}\boldsymbol{\mathrm{v}}_\ue= {\rm{Re}}\left[\mathcal{F}^{-1}\left(\mathcal{F}(\boldsymbol{\mathrm{C}}_0)\cdot\mathcal{F}(\boldsymbol{\mathrm{v}}_\ue)\right)\right],
\]
where $\mathcal{F}$ and $\mathcal{F}^{-1}$ are the FFT and anti-FFT operator, respectively, while $\cdot$ is the component wise product.
With the adoption of this approach, the computational time is noticeably reduced: each FFT computation requires $O(m\times \log_2(2m))$ operations. 
Finally, in order to compute equation~\eqref{eq:numerical_CK} we need to repeat the algorithm at each time step, and on the whole the computational burden 
can be estimated to be of order $O(n\times m\times \log_2 m)$, which is definitely a satisfactory improvement with respect to the non-FFT based procedure.

\subsection{Fast convolution at work: numerical result}\label{subsec:nr-mnoise}
Equipped with FCA we are now ready to approach multiplicative processes previously described. 
We want to check that numerical results obtained by fast convolution
converge to the analytical solution,  when available, or to the PDF reconstructed by means of MC simulation. 
Moreover, we show that FCA provides an estimate of the distribution shape even in those low probability regions, such as the tails,
which are inefficiently sampled by MC approach.

Lamperti transform for process~\eqref{eq:quadraticSDEtau} can be explicitly computed as
\begin{equation}\label{eq:quadraticLamperti}
	Z_\tau = \int_{X_0}^{X_\tau} \frac{\ud \hat{X}_\tau}{\sqrt{c\hat{X}_\tau^2+d\hat{X}_\tau+\tilde{e}(\tau)}}
	= \frac{1}{\sqrt{c}}\mathrm{asinh}\left(\frac{X_\tau+d/(2c)}{\sqrt{A^2_\tau}}\right)-\zeta_\tau^0\, , 
\end{equation}
with $A^2_\tau=(4\tilde{e}_\tau c-d^2)/(4c^2)$, and $\zeta_\tau^0=\mathrm{asinh}[(X_0+d/(2c))/\sqrt{A^2_\tau}]/\sqrt{c}$.\\
The related drift function is
\[
\mathrm{M}_Z(Z_\tau,\tau)=\frac{1}{\sqrt{c}}\left[\left(a-\frac{c}{2}\right)-\frac{\tilde{e}_{\tau}'}{2cA^2_\tau}\right]\tanh\left[\sqrt{c}(Z_\tau+\zeta_\tau^0)\right]
-\frac{\frac{d}{2c}\left(a-\frac{c}{2}\right)-b+\frac{d}{4}}{\sqrt{cA^2_\tau}\cosh\left[\sqrt{c}(Z_\tau+\zeta_\tau^0)\right]}
+\frac{\tilde{e}_{\tau}'}{2c^\frac{3}{2}A^2_\tau}\chi_\tau^0\, ,
\]
where $\chi_\tau^0=(X_0+d/(2c))/\sqrt{A^2_\tau+(X_0+d/(2c))^2}$, and the prime is a shorthand for the derivative w.r.t $\tau$. \\
\begin{figure}
	\begin{center}
		\subfloat[]{\resizebox*{0.5\textwidth}{!}{\includegraphics{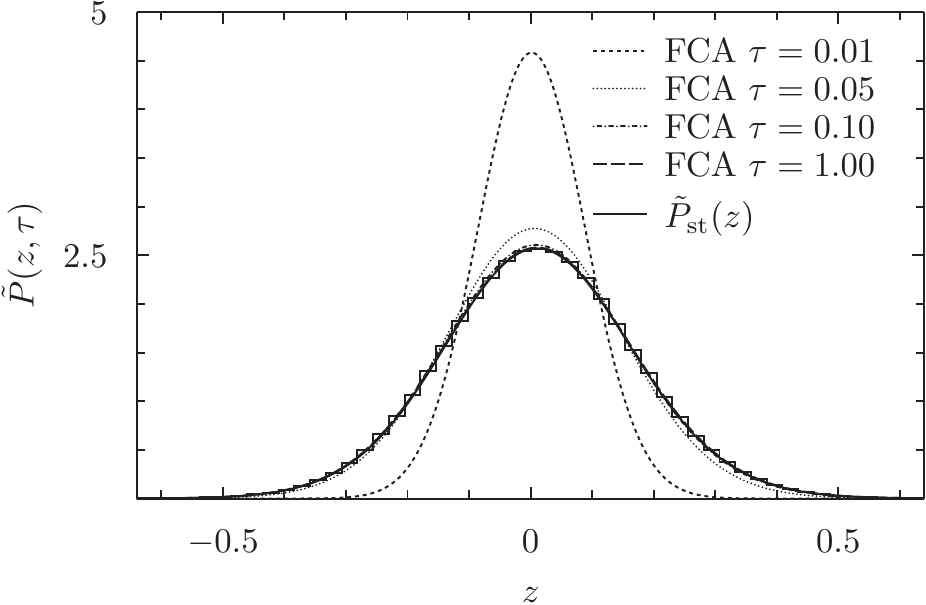}}\label{fig:const_scaling_panelA}}%
		\subfloat[]{\resizebox*{0.5\textwidth}{!}{\includegraphics{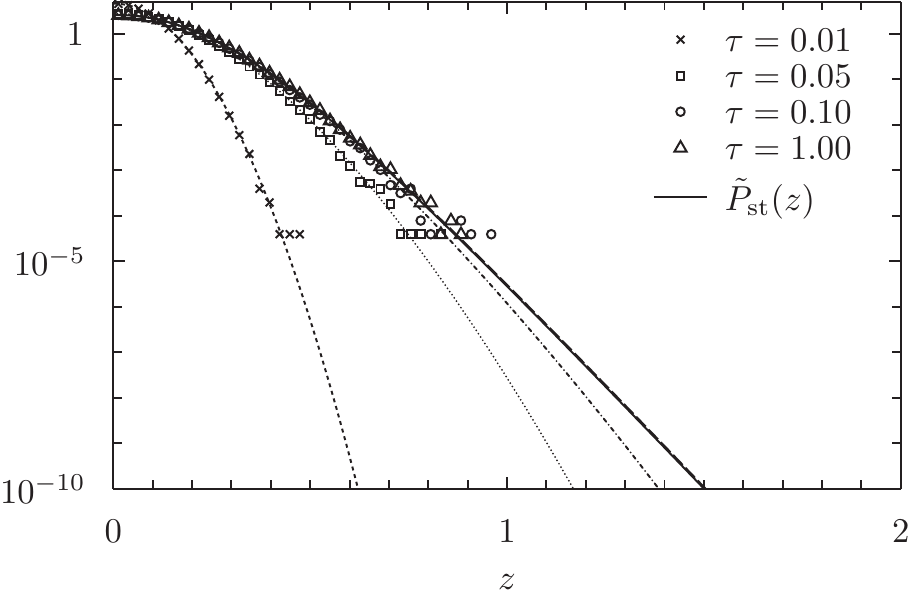}}\label{fig:const_scaling_panelB}}%
	\end{center}
	\caption{PDF of $Z_\tau$ for increasing values of $\tau$; solid line corresponds to the analytical stationary solution, dashed ones to FCA, 
	while bars in Panel~(a) and symbols in Panel~(b) to MC simulation of the process for maturity $\tau=1$.}
	\label{fig:const_scaling}
\end{figure}
The first and simplest case we want to consider corresponds to the SDE~\eqref{eq:quadraticSDEtau} with time independent 
parameter $e>0$, $\mathrm{D}^2>0$, and negative $a$. Whenever this conditions are satisfied, 
the process converges exponentially to the stationary regime with a typical relaxation time given by $-1/a$. 
Following~\cite{Biro_Jakovac} the stationary PDF can be computed in closed-form expression as
\begin{equation}\label{eq:stationary}
	P_\mathrm{st}(x)\propto\frac{1}
	{\left[\left(x+\frac{d}{2c}\right)^2+\frac{\mathrm{D}^2}{4c^2}\right]^{\frac{1+\nu}{2}}}
	\exp{\left[-2\frac{ad-2bc}{c\sqrt{\mathrm{D}^2}}\mathrm{atan}\left(\frac{x\sqrt{\mathrm{D}^2}}{2e+d x}\right)\right]}
	\, ,
\end{equation}
with $\nu=1-2a/c$, and the inverse tangent function continues smoothly at $x>-2e/d$. For illustrative purposes we fix the five free parameters as 
$a=-20$, $b=d=e=0.1$, and $c=4.5$, while the choice of $g(t)$ and $t_0$ is at the moment irrelevant since 
we work directly with time $\tau$. As evident from equation~\eqref{eq:stationary} all moments of order $n$ higher than or equal to $\nu$ diverge:
being $\nu\simeq 9.9$, only the first nine lowest moments converge. In figure~\ref{fig:const_scaling_panelA} we plot the time 
evolution of $\tilde{P}(z_\tau,\tau)$ for increasing values of $\tau=0.01,0.05,0.1,1$ as obtained by means of FCA 
($z_\mathrm{min}=-10.24$, $m=2^{13}$, $\Delta z=-2z_\mathrm{min}/m$, and $\Delta\tau=10^{-3}$). 
For $\tau=1$ we also plot the histogram corresponding to MC simulation of the discrete process
(parameter of the Euler scheme approximation: $\Delta\tau=10^{-3}$, and number of MC paths $N_\mathrm{MC}=10^6$), while the solid line represents the analytical solution easily 
derived from equation~\eqref{eq:stationary}. In figure~\ref{fig:const_scaling_panelB} we show the same results in log-linear scale to emphasize the tail region. 
The analytical information provides an overall check that the algorithm converges to the correct distribution, 
however far from stationary regime we have no precise information about the PDF shape. 
In~\cite{Bormetti_Delpini} the scaling of the convergent moments is computed analytically, but it is known that the knowledge of the moments does not allow for a
univocal reconstruction of the complete distribution. We can only rely on MC simulation, but 
sampling of low probability region requires on average a huge statistics (we need  $N_\mathrm{MC}>1/p$ to explore a $p$-probability region).
At this point the advantages provided by the fast convolution based approach are evident, as clearly shown by both panels.  
FCA curve for $\tau=1$ is in perfect agreement with the analytical prediction, both in central and tail regions. MC histograms agree as well, but
these results are extremely noisy and very inaccurate for $\tilde{P}(z_\tau,\tau)\lesssim 10^{-4}$.
FCA based results are even more impressive looking at the computational time. 
Performances are strongly machine dependent, and for this reason we do not quote absolute times, but measured relative values: 
to obtain $\tau=1$ bars MC takes ten times more than FCA\footnote{Random numbers generators and FFT algorithms are provided by GNU Scientific Library.}.
As a consequence it needs $10^7$ times more to reach the same accuracy at $\tilde{P}(z_\tau,\tau)\sim 10^{-10}$ level. 
\begin{figure}
	\begin{center}
		\subfloat[]{\resizebox*{0.5\textwidth}{!}{\includegraphics{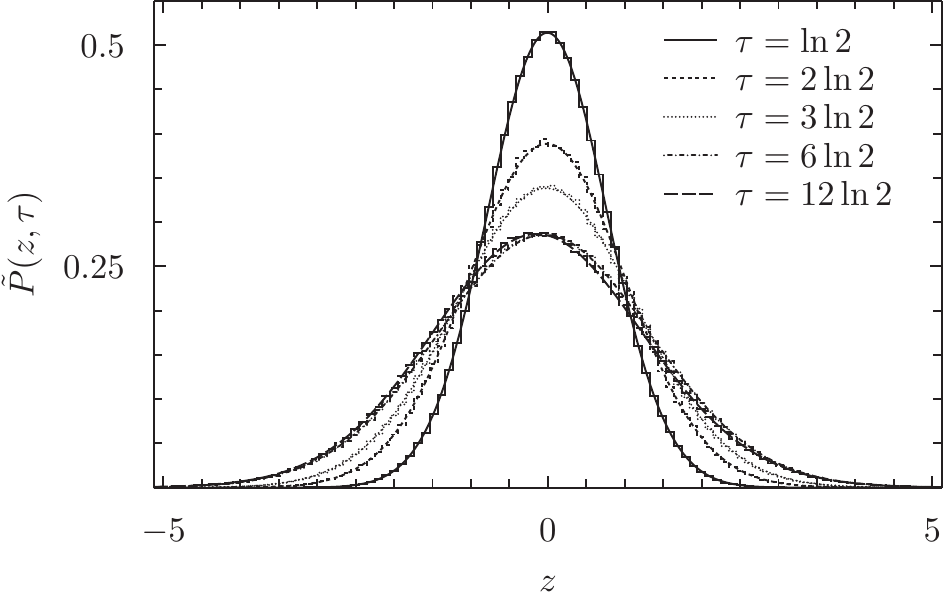}}\label{fig:etdep_scaling_panelA}}%
		\subfloat[]{\resizebox*{0.5\textwidth}{!}{\includegraphics{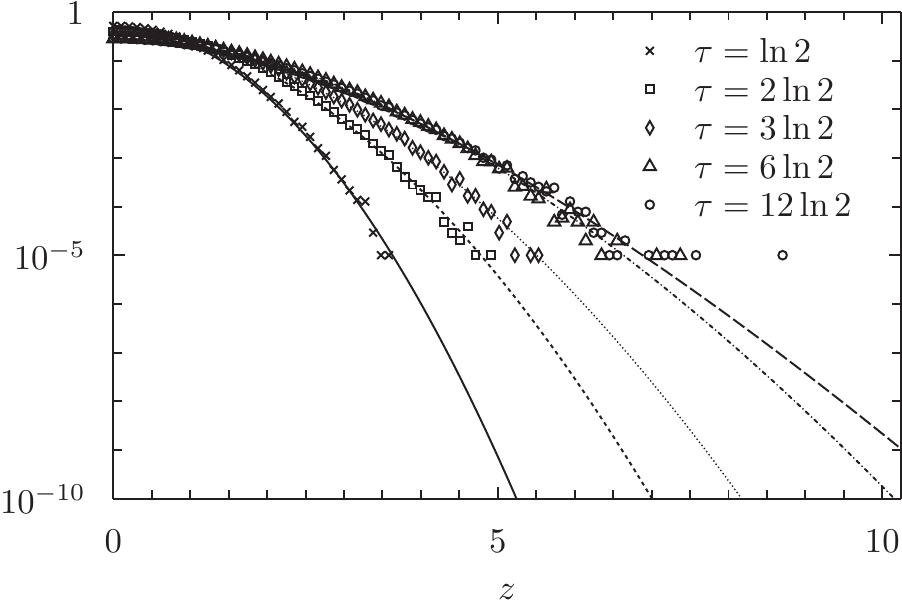}}\label{fig:etdep_scaling_panelB}}%
		\label{fig:etdep_scaling}
	\end{center}
	\caption{PDF of $Z_\tau$ for increasing values of $\tau$; lines correspond to FCA, while bars in Panel~(a) and symbols in Panel~(b) to MC simulations.}
\end{figure}
Similar results are obtained for the slightly more complicated process used in~\cite{Friedrich_etal} to model foreign exchange rate fluctuations.
Their process is still mean reverting with $a=-4.4\times 10^{-1}$, $b=0$, $c=3.8\times 10^{-2}$, and $d=3.04\times 10^{-3}$, though in this case
the last parameter has a non trivially time dependence, $\tilde{e}(\tau)=6.08\times 10^{-5}+6\times 10^{-3}\exp{(-0.5\tau)}$. Lacking stationarity,
every analytical information on the PDF is lost, yet we can see from figures~\ref{fig:etdep_scaling_panelA} and~\ref{fig:etdep_scaling_panelB} 
how the numerical PDF evolves with time and we verify a striking matching between MC and FCA results. Remarks similar to previous case apply. 

We now turn our attention to piecewise linear diffusion. The procedure is in this case a little bit subtle. 
While computation of the Lamperti transform of process~\eqref{eq:piecewiseSDE} 
for $H=1/2$ and integral time $\tau=2\sqrt{t}$ is still feasible providing~\footnote{The sign function is defined according to the convention $\mathrm{sign}(0)=0$.}
\[
Z_\tau = \frac{2}{\sigma\epsilon}\mathrm{sign}(X_\tau)\left(\sqrt{\frac{\tau}{2}+\epsilon\abs{X_\tau}}-\sqrt{\frac{\tau}{2}}\right)\, ,
\]
the stochastic differential $\ud Z_\tau$ cannot be computed applying It\^o Lemma straightforwardly. As a function of $X_\tau$ and $\tau$, 
$Z_\tau$ lacks necessary regularity condition for $\tau=0$ and $X_\tau=0$. However, both difficulties can be overcome. 
The $X_\tau$ process does not suffer any problem in $\tau=0$, therefore we can evolve from $X^0$ to $X^1$, and then exploit the one to one correspondence between
$X_\tau$ and $Z_\tau$. Moving from $\tau=0$ to $\tau=\Delta \tau$, $X^1$ remains delta distributed around zero, and the same holds true for $Z^1$. 
For $\tau\geq\Delta\tau$ the time derivative $\partial Z_\tau/\partial\tau$ needed in $\ud Z_\tau$ can be readily computed.
The difficulty arising with the computation of $\partial^2 Z_\tau/\partial X_\tau^2$ in zero can be dealt by
replacing the absolute value with the smooth approximation
\[
\abs{X_\tau}_\mathrm{s}\doteq X_\tau\left(\frac{2}{1+\ue^{-2kX_\tau}}-1\right)\, .
\]
This allows us to compute 
\[
\frac{\ud}{\ud X_\tau}\abs{X_\tau}_\mathrm{s}=\left(\frac{2}{1+\ue^{-2kX_\tau}}-1\right)+4kX_\tau\frac{\ue^{-2kX_\tau}}{\left(1+\ue^{-2kX_\tau}\right)^2}
\simeq\mathrm{sign}(X_\tau)\, ,
\]
where in view of our application on a discrete grid the last approximation can be justified for sufficiently large $k$.
\begin{figure}
	\begin{center}
		\subfloat[]{\resizebox*{0.5\textwidth}{!}{\includegraphics{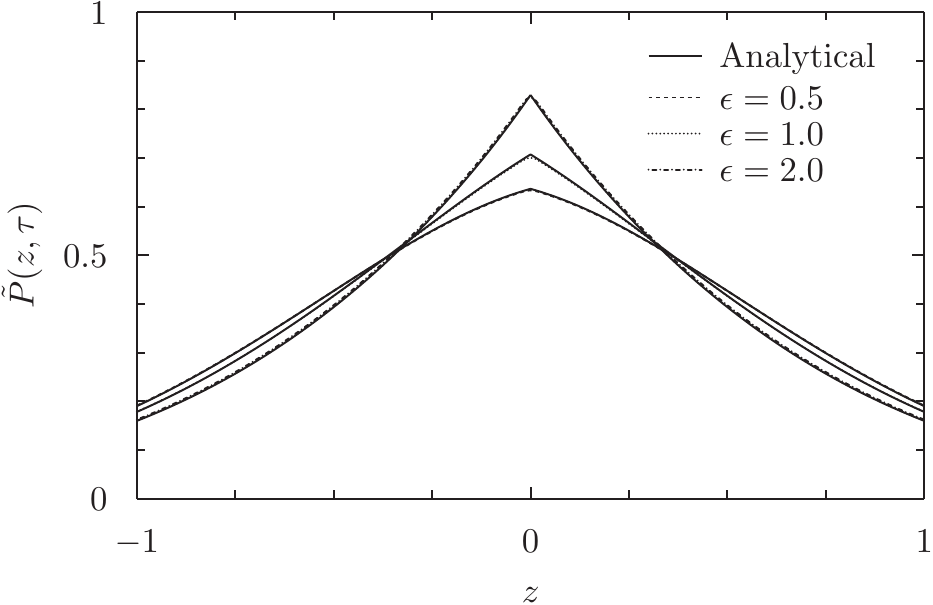}}\label{fig:piecewise_panelA}}%
		\subfloat[]{\resizebox*{0.5\textwidth}{!}{\includegraphics{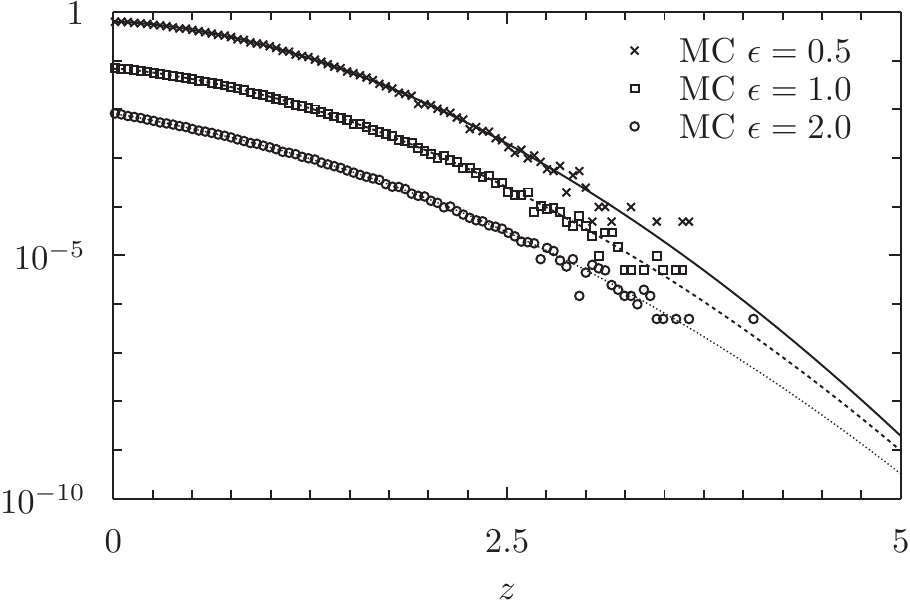}}\label{fig:piecewise_panelB}}%
		\label{fig:piecewise}
	\end{center}
	\caption{PDF of $Z_\tau$ at time $\tau=1$ for $\sigma^2=1$ and $\epsilon=0.5, 1, 2$. 
	Panel (a): comparison between analytical expressions (solid lines) and FCA (dashed and dotted lines);
	Panel (b): comparison between MC histograms (symbols) and FCA. Log-linear curves have been shifted for readability.}
\end{figure}
We thus end with the following expression for the dynamics of $Z_{\tau}$
\[
\ud Z_\tau\simeq\mathrm{sign}(Z_\tau)\left[\frac{1}{2\epsilon}\left(\frac{1}{\frac{\sigma^2\epsilon}{2}\abs{Z_\tau}+\sigma\sqrt{\frac{\tau}{2}}}
-\frac{1}{\sigma\sqrt{\frac{\tau}{2}}}\right)-\frac{\epsilon\sigma^2}{4}\frac{1}{\frac{\sigma^2\epsilon}{2}\abs{Z_\tau}+\sigma\sqrt{\frac{\tau}{2}}}\right]\ud\tau + \ud W_\tau\, .
\]
Numerical results concerning this last process are reported in figure~\ref{fig:piecewise_panelA} (linear scale) 
and in figure~\ref{fig:piecewise_panelB} (log-linear scale).
We study the dependence of $\tilde{P}(z,\tau)$ on $1/(\sigma^2\epsilon)$. 
Indeed for $\abs{x}\gg 1$,  $P(x,\tau)\sim \exp{[-2\abs{x}/(\sigma^2\epsilon\tau)]}$, 
and the value of the coefficient in the exponential function is crucial to asses the convergence of the expectation of $\exp(x)$ 
with respect to $P(x,\tau)$. We fix $\tau=1$, $\sigma^2=1$, and $\epsilon = 0.5, 1, 2$. 
The leptokurtosis of the PDF increases as far as the value of $\epsilon$ increases. Parameters for the Euler scheme approximation are fixed 
as in previous examples, while for the FCA we have slightly changed the value of $\Delta\tau=10^{-4}$ and $m=2^{11}$ keeping $z_\mathrm{min}=-10.24$.
For each one of the three cases we also plot the analytical prediction, since for piecewise linear processes the solution is known in closed form. 
Also in this last case the agreement between analytical and fast convolution PDF is totally satisfactory, 
while limitations of the MC approach are evident from symbols depicted in Panel~(b).

\section{Financial applications}\label{sec:fin_application}
In the second part of this work we present and discuss how results achieved in previous sessions 
can be exploited in finance, and in particular in the context of option pricing.
For both quadratic and piecewise diffusion we briefly review how to set the correct risk neutral framework. 
Then, for explanatory purposes, we apply FCA to price European Plain Vanilla and geometric Asian options, 
but the approach can be extended to deal with different payoffs and different kind of boundary conditions.    
For the remaining of this paper, $X_t=\ln{S_t}-\ln{S_{t_0}}$ is the logarithmic return obtained from the stochastic 
process $S_t$ describing the evolution of an asset price. As asset candidates we only consider equities and 
foreign exchange rates. 

\subsection{Piecewise diffusion under risk-neutrality: Plain Vanilla pricing}\label{subsec:riskneutral}
According to risk neutral valuation theory we need to find the dynamics of $S_t$ or, equivalently, $X_t$ under the 
probability measure which makes all discounted asset prices martingales. Whenever the Novikov condition 
for the process under consideration is verified, Girsanov theorem gives the recipe for the equivalent measure, 
and it also explains how the dynamics of $S_t$ coherently modifies. However,~\cite{McCauley_Gunaratne,McCauley_etal.2007b} show 
how to compute the desired martingale directly from the Green function solving the FP equation associated to the dynamics
\begin{equation}\label{eq:dynamicsSt}
	\ud S_t = \mu S_t\ud t + \sigma S_t\sqrt{1+\epsilon\frac{\abs{\ln S_t-\ln S_{t_0}}}{\sqrt{t}}}\ud W_t\, ,
\end{equation}
with $S_{t_0}=S_0$. Just in the case of the original model of~\cite{Black_Scholes,Merton}, a delta hedged strategy allows 
to construct a locally risk neutral portfolio and to derive the partial differential equation 
\begin{equation}\label{eq:BSMgeneral}
	\frac{\partial O}{\partial t} + rS_t\frac{\partial O}{\partial S_t}
	+\sigma^2 S_t^2 \frac{\sqrt{t}+\epsilon\abs{\ln{S_t}-\ln{S_0}}}{2\sqrt{t}}\frac{\partial^2 O}{\partial S_t^2}-rO=0\, ,
\end{equation}
that is to be use to solve the pricing problem of a Plain Vanilla option $O$, with $r$ risk free interest rate and for suitable boundary conditions.
Introducing $\hat{O}(S_t,t)\doteq\ue^{r(T-t)}O(S_t,t)$ and substituting in equation~\eqref{eq:BSMgeneral}, 
it is readily verified that the hat price satisfies an equation formally identical to the backward time FP equation 
associated with the dynamics~\eqref{eq:dynamicsSt} with $\mu=r$. The fair price of a call option is therefore predicted to be
\[
O(S_0,t_0)=\ue^{-r(T-t_0)}\int_{-\infty}^{+\infty}\ud S_T \left(S_T-K\right)^+\mathrm{G}^\mathbb{Q}(S_T,T;S_0,t_0)
=\ue^{-r(T-t_0)}\mathbb{E}^\mathbb{Q}\left[(S_T-K)^+\rvert S_0\right]\, ,
\]
where $\mathrm{G}^\mathbb{Q}$ is the Green function solving the FP equation in the risk neutral framework, and $K$ is the strike price. 
The dynamics of $X_t$ under the new probability measure reads 
\begin{equation}\label{eq:Xtriskneutral}
	\ud X_t = \left[r-\frac{\sigma^2}{2}\left(1+\epsilon\frac{\abs{X_t}}{\sqrt{t}}\right)\right]\ud t+\sigma \sqrt{1+\epsilon\frac{\abs{X_t}}{\sqrt{t}}}\ud W_t^\mathbb{Q}\, ,
	\quad X_{t_0}=0\, .
\end{equation}
At variance with equation~\eqref{eq:piecewiseSDE}, a non trivial drift term appears and some comments are mandatory.
As recognized by McCauley and collaborators, whenever the drift depends explicitly
on $X_t$ there is no way to preserve scaling properties. However, in order to exploit the analytical information provided by 
equation~\eqref{eq:exponential}, corresponding to the Green function $\mathrm{G}^\mathbb{Q}(X_T,T;0,0)$ for the process~\eqref{eq:piecewiseSDE}, 
they provide arguable arguments in order to replace the drift with a constant. 
We are instead equipped with a computationally efficient algorithm, and so we can get rid of this approximation and price options directly 
with the process~\eqref{eq:Xtriskneutral}. 
As in section~\ref{sec:mnoise}, we switch to integral time
\[
	\ud X_\tau=\left(r-\frac{\sigma^2}{2}\right)\frac{\tau}{2}\ud\tau-\frac{\epsilon\sigma^2}{2}\abs{X_\tau}\ud\tau+\sigma\sqrt{\frac{\tau}{2}
	+\epsilon\abs{X_\tau}}\ud W_\tau^\mathbb{Q}\, ,\quad X_0=0\, ,
\]
and compute the Call option price as
\begin{align}
	O(S_0,t_0)&=S_0^D\mathbb{E}^\mathbb{Q}\left[(\ue^{X_{\tau(T)}}-\ue^\mathrm{k})^+\rvert X_0\right]
	=S_0^D\mathbb{E}^\mathbb{Q}\left[(\ue^{X(Z_{\tau(T)})}-\ue^\mathrm{k})^+\rvert Z_0\right]\nonumber\\
	&=S_0^D\int_{-\infty}^{+\infty}\ud z_{\tau(T)}
	\left[\ue^{\mathrm{sign}(z_{\tau(T)})\frac{1}{\epsilon}\left[\left(\frac{\sigma\epsilon}{2}\abs{z_{\tau(T)}}+\sqrt{\frac{\tau(T)}{2}}\right)^2
	-\frac{\tau(T)}{2}\right]}-\ue^\mathrm{k}\right]^+p^\mathbb{Q}(z_{\tau(T)}\rvert z_0)\nonumber\\
	&\simeq S_0^D\Delta z\sum_{j=0}^{m-1}\left[\ue^{\mathrm{sign}(z_\mathrm{min}+j\Delta z)\frac{1}{\epsilon}
	\left[\left(\frac{\sigma\epsilon}{2}\abs{z_\mathrm{min}+j\Delta z}+\sqrt{\frac{n\Delta\tau}{2}}\right)^2
	-\frac{n\Delta\tau}{2}\right]}-\ue^\mathrm{k}\right]^+{\mathrm{P}_j^n}^\mathbb{Q}\label{eq:callprice}\, ,
\end{align}
with $\tau(T)=2\sqrt{T}$, discounted price $S_0^D=\ue^{-r(T-t_0)}S_0$ and log-moneyness $\mathrm{k}=\ln{(K/S_0)}$. 
The vector ${\mathrm{P}_j^n}^\mathbb{Q}$ of transition probability between $z_0$ and $z_j^n$ under the risk neutral measure $\mathbb{Q}$ has to be 
computed with the fast convolution procedure described in section~\ref{subsec:FCA}. 

\subsection{Exotic options: the geometric Asian case}\label{subsec:asian}
Formula~\eqref{eq:callprice} can be extended to deal with payoffs with different dependence on $S_\tau$, e.g. digital options, 
covered call or strongly non linear function $f(S_\tau)$, the only constraint being $\mathbb{E}^\mathbb{Q}[f(S_\tau)\rvert S_0]<\infty$. 
The case of a functional payoff depending multiplicatively on the price along the path, i.e. 
$f([S_\tau])=\prod_{i=0}^{n}f_i(S^i)$, is just slightly more complicated but in fact can be easily managed,
see~\cite{Chiarella_ElHassan} for an application to bond pricing. 
In this section we address the problem of pricing a geometric Asian option, which requires to compute the  expected value
\begin{equation}\label{eq:geoAsian}
	\mathbb{E}^\mathbb{Q}\left[\left(\ue^{\frac{1}{T-t_0}\int_{t_0}^T \ln{S_s}\ud s}-K\right)^+ | S_0\right]\, .
\end{equation}
Being the positive part function non linear, previous expression is quite tricky to be evaluated and requires some careful manipulations.  
Defining $\tau=2(\sqrt{t}-\sqrt{t_0})$, the Asian price is given by
\begin{equation}
	O_\mathrm{A}(S_0,t_0)=S_0^D\mathbb{E}^\mathbb{Q}\left[\left(\ue^{\frac{1}{T-t_0}
	\int_{0}^{\tau(T)} \left(\frac{\tau'}{2}+\sqrt{t_0}\right)X_{\tau'}\ud\tau'}-\ue^\mathrm{k}\right)^+ | X_0\right]\, .
\end{equation}
We exploit the discretization of the $\tau(T)$ time interval in $n$ equally spaced intervals of amplitude $\Delta\tau$, and we replace 
the integral expression with a finite sum $U^n\doteq\sum_{j=1}^{n}\left(j\Delta\tau/2+\sqrt{t_0}\right)x^{j}/n$.
We then introduce the ancillary variables $\{U^1,\ldots,U^n\}$ satisfying the following recursive relation
\begin{equation}\label{eq:U}
	U^{i+1}= \frac{i}{i+1} U^i + \left(\frac{\Delta\tau}{2}+\frac{\sqrt{t_0}}{i+1}\right)X^{i+1}\, ,
\end{equation}
for $i=1,\ldots,n-1$ and $U^1=\left(\frac{\Delta\tau}{2}+\sqrt{t_0}\right)X^1$.
Exploiting the one to one correspondence between $X_\tau$ and $Z_\tau$, it is possible to rewrite the Asian price as
\[
O_\mathrm{A}(S_0,t_0)=S_0^D\int_{u^n}\ud u^n\mathrm{A}(u^n)p_U^\mathbb{Q}(u^n)
                     =S_0^D\int_{u^n}\ud u^n\int_{z^n}\ud z^n\mathrm{A}(u^n)p_{U Z}^\mathbb{Q}(u^n,z^n)\, ,
\]
with $\mathrm{A}(u^n)=\left(\ue^{2\frac{\sqrt{T}-\sqrt{t_0}}{T-t_0}u^n}-\ue^\mathrm{k}\right)^+$. 
The only unknown quantity in previous expression is the joint distribution of $U^n$ and $Z^n$, whose computation  
requires a recursive relation allowing to propagate $p_{U Z}^\mathbb{Q}(u^i,z^i)$ to $p_{U Z}^\mathbb{Q}(u^{i+1},z^{i+1})$
with the associated initial time condition $p_{U Z}^\mathbb{Q}(u^1,z^1)=\delta(z^1)\delta(u^1)$.
The following equation holds
\begin{equation}\label{eq:UZtoUZ}
	p_{U Z}^\mathbb{Q}(u^{i},z^{i+1}) = \int_{z^i}\ud z^i p_{U Z}^\mathbb{Q}(u^i,z^i) \pi^\mathbb{Q}(z^{i+1}\rvert z^i)\, ,
\end{equation}
and to proceed it is useful to explicit the dependence of $X^{i+1}$ on $Z^{i+1}$ in equation~(\ref{eq:U})
\begin{equation}
\left\{
	\begin{array}{l} 
		U^{i+1}=\frac{i}{i+1}U^{i}+\left(\frac{\Delta\tau}{2}+\frac{\sqrt{t_0}}{i+1}\right)\mathrm{sign}(Z^{i+1})\frac{1}{\epsilon}
		\left[\left(\frac{\sigma\epsilon\abs{Z^{i+1}}}{2}
		+\sqrt{\frac{\tau}{2}}\right)^2-\frac{\tau}{2}\right]\\
		Z^{i+1}=Z^{i+1}\, ,
	\end{array}
	\right.\label{eq:UZsystem}
\end{equation}
where $U^{i+1}$ is coupled with the dummy variable $Z^{i+1}$. From previous relations we have
\begin{equation}\label{eq:Jacobian}
	p_{U Z}^\mathbb{Q}(u^{i+1},z^{i+1}) = \abs{\frac{\partial(u^{i},z^{i+1})}{\partial (u^{i+1},z^{i+1})}}p_{U Z}^\mathbb{Q}(u^{i}(u^{i+1},z^{i+1}),z^{i+1})\,,
\end{equation}
where the Jacobian is equal to $(i+1)/i$. 
Therefore starting from the distribution $p_{U Z}^\mathbb{Q}(u^1,z^1)$, and following the above procedure, after $n-1$ steps we obtain 
the desired $p_{UZ}^\mathbb{Q}(u^n, z^n)$.
Introducing $m_Z$-node grid for $Z^i$ and $m_U$-node grid for $U^i$, we can approximate the distribution $p_{UZ}^\mathbb{Q}(u^i, z^i)$  
with a $m_U\times m_Z$ matrix ${\mathrm{P}^i_{jk}}^\mathbb{Q}$, the row index $j$ running over the nodes of $U^i$, the column index $k$ over those of $Z^i$.
The Asian price can therefore be approximated by
\[
O_\mathrm{A}(S_0,t_0)\simeq S_0^D\Delta u\Delta z\sum_{j=0}^{m_U-1}\mathrm{A}(u^n_j)\sum_{k=0}^{m_Z-1}{\mathrm{P}^i_{jk}}^\mathbb{Q}\,.
\]
Since the time evolution corresponding to equation~\eqref{eq:UZtoUZ} is the most computationally intensive operation implicit in previous approximation, 
we can perform it at each node $u^i_j$ by means of FCA. The overall numerical complexity of the algorithm is essentially linear in 
the total number of grid nodes, i.e $O(n\times m_U\times m_Z\log_2{m_Z})$.

\subsection{The Vellekoop-Nieuwenhuis-Borland model}\label{subsec:VNB_model}
The geometric Asian case just described is useful also in view of the last application we present, 
which is related to the model for the stock price dynamics introduced in~\cite{Borland-PRL,Borland-QF}.
Borland model tries to generalize the standard Black\&Scholes to account for the empirical evidences of fat tailed return distributions, 
still keeping a closed form formula for the price of Plain Vanilla instruments. 
It is a sort of hybrid between a stochastic volatility model and the standard  Black\&Scholes: the volatility is stochastic, but the stock price 
and the volatility itself are driven by the same Brownian motion. 
Despite its theoretical elegance, the model contains some weaknesses extensively analyzed in~\cite{Vellekoop-Nieuwenhuis}.
They raised two main objections by proving the Borland model to suffer arbitrage opportunities and diverging payoff expectation, 
as a consequence of the thickness of the tails. However, they preserved the main idea of the model and they amended it from both drawbacks. 
In their modified version they introduce the following dynamics 
\begin{align}
	\ud S_t&=\mu S_t\ud t+\sigma S_t \ud\Omega_t\, ,\quad S_{t_0}=S_0\, ,\label{eq:objmeas_vn}\\
	\ud\Omega_t&=\Sigma(\Omega_t,t)\ud W_t\, ,\quad \Omega_{t_0}=\Omega_0\, ,\label{eq:omegas}
\end{align}
with
\begin{equation}\label{eq:Sigma}
	\Sigma(\Omega_t,t)=\left\{
	\begin{array}{ccc}
		A^{-\frac{\alpha}{2}}\mathrm{P}(\Omega_t,t)^{-\frac{\alpha}{2}}&\quad &t>0\\
		0&\quad &t=0\, ,
	\end{array}
	\right.
	\quad\mathrm{and}\quad
	\mathrm{P}(\Omega_t,t)=\frac{1}{N_t}(1+\alpha\beta_t\Omega_t^2)^{-\frac{1}{\alpha}}\, ,
\end{equation}
with $\beta_t=\left[(1-\alpha)(2-\alpha)t\right]^{-\frac{2}{2-\alpha}}$, $N_t=A/\sqrt{\beta_t}$, 
$A=\sqrt{\frac{\pi}{\alpha}}\Gamma\left(\frac{1}{\alpha}-\frac{1}{2}\right)/\Gamma\left(\frac{1}{\alpha}\right)$,
$\alpha\in\left(0, \frac{1}{2}\right)$ and $t_0\geq 0$.
In~\cite{Vellekoop-Nieuwenhuis} the existence of a solution for equation~\eqref{eq:omegas} is proved, and it is shown how the unconditional distribution 
(i.e. $\Omega_{t_0}=0$ for $t_0=0$) reduces to the generalized Student-$t$ distribution. In general the conditional distribution deviates from it.
The log-return $X_t$ satisfies the equation
\[
 X_T=X_t+\mu (T-t)-\frac{1}{2}\sigma^2\int_t^T \Sigma^2(s, \Omega_s)\ud s+ \sigma(\Omega_T-\Omega_t)\, ,
\]
for $T>t\geq t_0$. They verify that sufficient conditions hold for the applicability of the Girsanov theorem, and
they derive the risk neutral dynamics
\begin{equation}\label{eq:dSdOriskneutral}
	\ud S_t=r S_t\ud t+\sigma S_t\ud\Omega_t^{\mathbb{Q}}\, ,\quad\mathrm{with}\quad\ud\Omega_t^\mathbb{Q}=\Sigma(\Omega_t,t)\ud W_t^{\mathbb{Q}}\, .
\end{equation}
This last process does not suffer anymore of previous problems, however $S_t$ does not satisfy the Markov property, yet only jointly with $\Omega_t$. 
In addition, the price of Plain Vanilla instruments cannot be given in closed form formula.
Indeed, according to pricing theory we have
\[
	O_\mathrm{C}(S_0,\Omega_0,t_0)=S_0^D\mathbb{E}^\mathbb{Q}\left[\left(\ue^{r(T-t_0)+\sigma(\Omega_T-\Omega_0)-\frac{1}{2}\sigma^2\int_{t_0}^T
	\Sigma^2(\Omega_s,s)\ud s}-\ue^\mathrm{k}\right)^+\Big\rvert S_0, \Omega_0\right]\, ,
\]
and the expectation can only be computed via numerical techniques.\\ 
Given~\eqref{eq:Sigma}, we observe that the equation governing the evolution of $\ud\Omega_t^\mathbb{Q}$ belongs to the class of quadratic diffusion 
processes~\eqref{eq:quadraticSDE} through the identifications $a=b=d=0$, $c=\alpha/\left[\left(1-\alpha\right)\left(2-\alpha\right)\right]$, 
$e(t)=\left[\left(1-\alpha\right)\left(2-\alpha\right)\right]^{\frac{\alpha}{2-\alpha}}t^{2/(2-\alpha)}$, and $g(t)=t$.
Switching to the integral time $\tau=\ln t/t_0$ for $t_0>0$, and recalling equation~\eqref{eq:quadraticLamperti}, $Z_\tau$ is readily computed
\begin{equation}\label{eq:ZfunctOmega}
Z_\tau=\frac{1}{\sqrt{c}}\left[\mathrm{asinh}\left(C_{\alpha,t_0,\tau}\Omega_\tau\right)-\mathrm{asinh}\left(C_{\alpha,t_0,\tau}\Omega_0\right)\right]\, ,
\end{equation}
where $C_{\alpha,t_0,\tau}=\sqrt{\alpha\beta_{t_0}}~\ue^{-\tau/(2-\alpha)}$. The price of a Plain Vanilla instrument can be computed as
\begin{equation}\label{eq:callVN}
	O_\mathrm{C}(S_0,Z_0,t_0)=S_0^D\mathbb{E}^\mathbb{Q}\left[\left(\ue^{r(T-t_0)+\sigma[\Omega(Z_{\tau(T)})-\Omega(Z_0)]
	-\frac{(2-\alpha)\sigma^2}{4}\left[e(T)-e(t_0)\right]-\frac{c\sigma^2}{2}\int_{0}^{\tau(T)}\Omega(Z_{\tau'})^2\ud \tau'}
	-\ue^\mathrm{k}\right)^+\Big\rvert S_0, Z_0\right]\, \end{equation}
with $\tau(T)=\ln T - \ln t_0$.
Defining the set of ancillary variables 
$\{U^1,\ldots,U^n\}$ satisfing the recursive relation
\begin{equation}\label{eq:UVN}
	U^{i+1}= U^i + \Delta\tau\Omega(Z^{i+1})^2\, ,
\end{equation}
with $U^1=\Delta\tau \Omega(Z^1)^2$, the formal analogy with the Asian case discussed in previous section is evident.
Computation of the expectation in~\eqref{eq:callVN} requires estimation of the joint probability $p_{U Z}^\mathbb{Q}(u^n,z^n)$. The procedure 
is identical to the Asian case; equation~\eqref{eq:UZtoUZ} is still valid, while the system~\eqref{eq:UZsystem} has to be coherently modified in 
\[
\left\{
	\begin{array}{l} 
		U^{i+1}=U^{i}+\frac{\Delta\tau}{C_{\alpha,t_0,(i+1)\Delta\tau}^2}\sinh^2\left[\sqrt{c}Z^{i+1}
		+\mathrm{asinh}\left(C_{\alpha,t_0,(i+1)\Delta\tau}\Omega(Z_0)\right)\right]\\
		Z^{i+1}=Z^{i+1}\, .
	\end{array}
	\right.
\]
The Jacobian in equation~\eqref{eq:Jacobian} simplifies to one, and, eventually, we can approximate the Plain Vanilla price as
\[
	O_\mathrm{C}(S_0,Z_0,t_0)\simeq S_0^D\Delta u\Delta z\sum_{j=0}^{m_U-1}\sum_{k=0}^{m_Z-1}\mathrm{C}(u^n_j,z_k^n){\mathrm{P}^i_{jk}}^\mathbb{Q}\, ,
\]
where $\mathrm{C}(u^n_j,z_k^n)=\left(\ue^{r(T-t_0)-\frac{\sigma^2}{2}\int_{t_0}^{T} \beta_s^{-\alpha/2}\ud s
+\sigma[\Omega(z_k^n)-\Omega(z_0)]-\frac{\alpha\sigma^2}{2(1-\alpha)(2-\alpha)}u_j^n}-\ue^\mathrm{k}\right)^+$,
and compute it by means of fast convolution. 

\subsection{Numerical results}\label{subsec:nr_plainv}
\begin{figure}
	\begin{center}
		\subfloat[]{\resizebox*{0.5\textwidth}{!}{\includegraphics{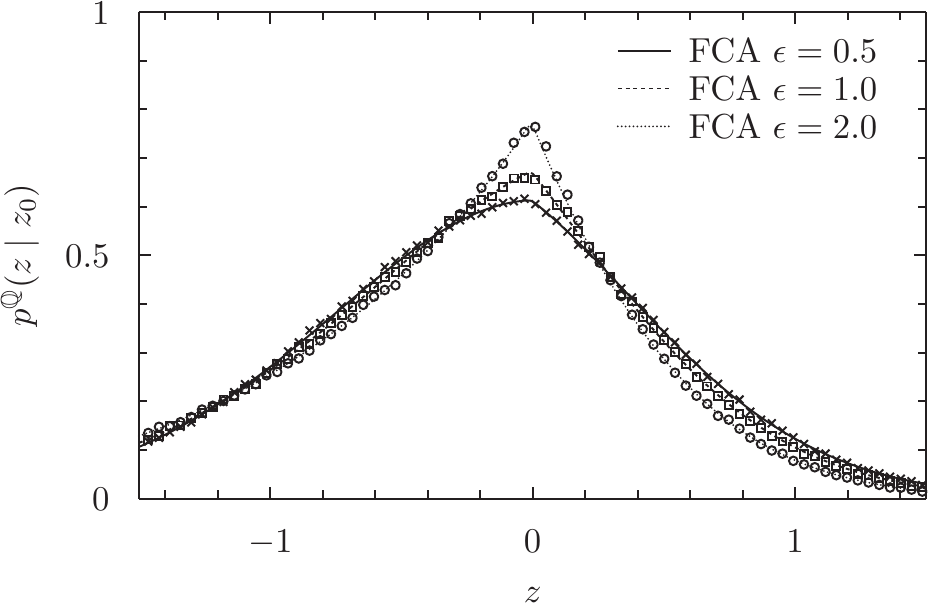}}\label{fig:RN_Piecewise_lin}}%
		\subfloat[]{\resizebox*{0.5\textwidth}{!}{\includegraphics{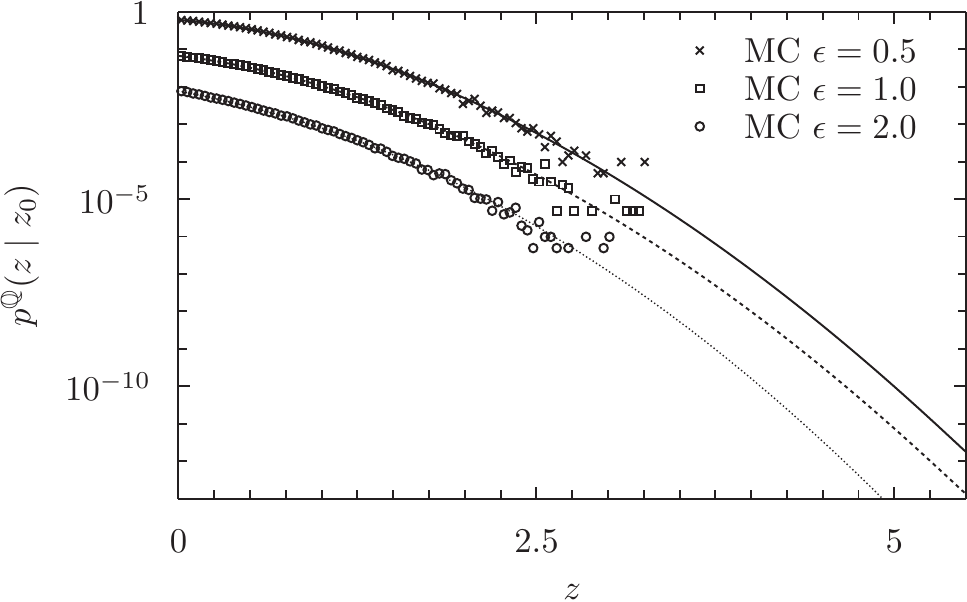}}\label{fig:RN_Piecewise_log}}%
		\label{fig:RN_Piecewise}
	\end{center}
	\caption{Piecewise diffusion: Risk neutral PDF of $Z_\tau$ at time $\tau=1$ for $r=0.03$, $\sigma^2=1$, and $\epsilon=0.5, 1, 2$. 
	Comparison between Monte Carlo histograms (symbols) and FCA (dashed and dotted lines); in Panel (b) curves have been shifted for readability.
	}
\end{figure}
\begin{figure}
	\begin{center}
		\subfloat[]{\resizebox*{0.5\textwidth}{!}{\includegraphics{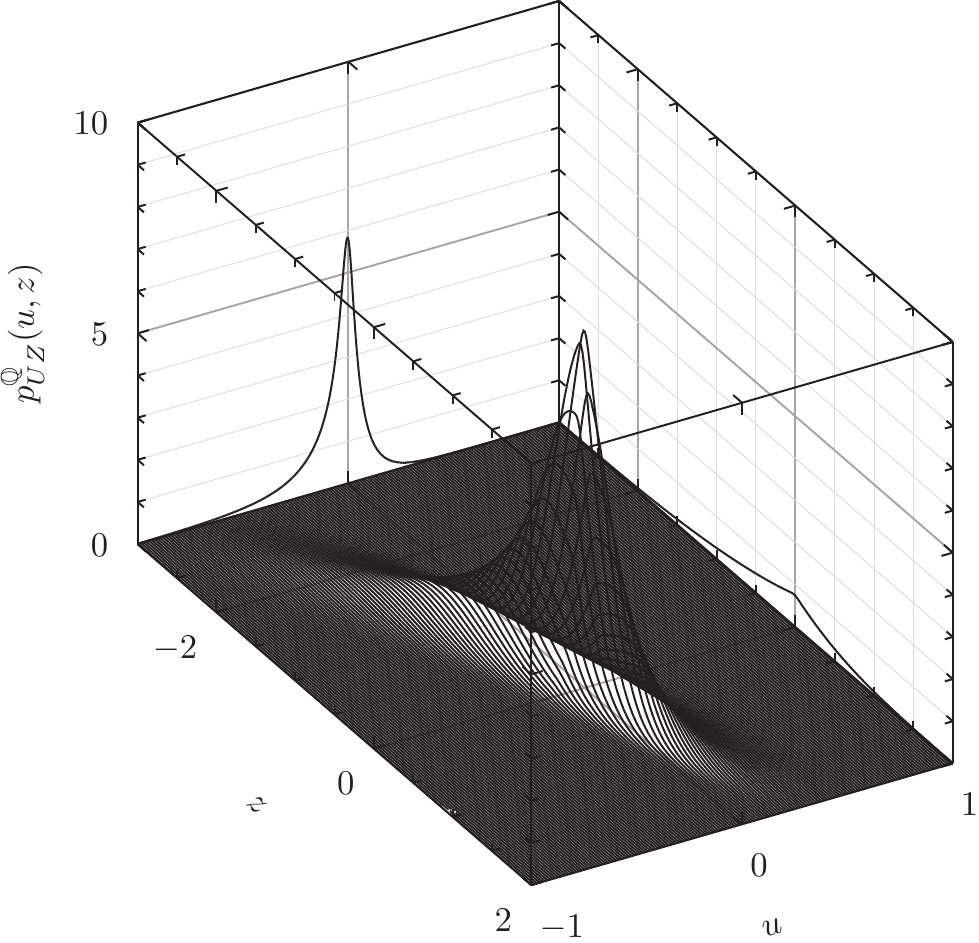}}\label{fig:2DPlot_Asian}}%
		\subfloat[]{\resizebox*{0.5\textwidth}{!}{\includegraphics{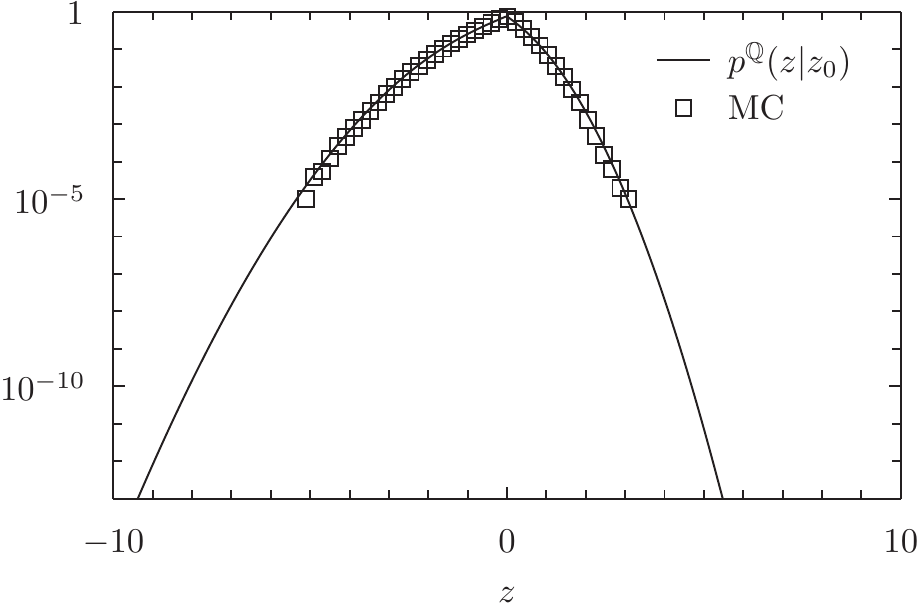}}\label{fig:RN-Asian_log}}%
		\label{fig:RN_Asian}
	\end{center}
	\caption{Piecewise diffusion: Bivariate risk neutral PDF of $Z_\tau$ and $U_\tau$, 
	and their corresponding marginals; $\epsilon=2$, $\sigma^2=1$, $t_0=0$, and $\tau=1$. In Panel (b)
	comparison between Fast Convolution PDF of Z and MC histogram.}
\end{figure}
In this final section we sum up numerical results for the financial applications described in paragraphs 
\ref{subsec:riskneutral}, \ref{subsec:asian}, and \ref{subsec:VNB_model}.\\
Whenever we switch to the risk neutral measure for the piecewise linear process,  corrections terms in the SDE appear and an 
analytical expression for the density is not available anymore. 
Numerical simulation is mandatory, and the FCA algorithm, being both faster and much more efficient, is a natural competitor to MC approach.
The transformed $Z_\tau$ process is enriched by the risk neutral correction (the last but one term in squared brackets)
\begin{align}
	\ud Z_\tau\simeq\mathrm{sign}(Z_\tau)\left[\frac{1}{2\epsilon}\left(\frac{1}{\frac{\sigma^2\epsilon}{2}\abs{Z_\tau}+\sigma\sqrt{\frac{\tau}{2}}}
	-\frac{1}{\sigma\sqrt{\frac{\tau}{2}}}\right)-\frac{\epsilon\sigma^2}{4}\frac{1}{\frac{\sigma^2\epsilon}{2}\abs{Z_\tau}+\sigma\sqrt{\frac{\tau}{2}}}\nonumber\right.\\
	\left.-\frac{1}{2}\left(\frac{\sigma^2 \epsilon}{2}\left|Z_{\tau}\right|+\sigma\sqrt{\frac{\tau}{2}}\right)
	+\frac{r}{\frac{\sigma^2 \epsilon}{2}|Z_\tau|+\sigma\sqrt{\frac{\tau}{2}}}
	\right]\ud\tau + \ud W_\tau\, ,\nonumber
\end{align}
where $r$ is the risk free rate. In figures~\ref{fig:RN_Piecewise_lin} and~\ref{fig:RN_Piecewise_log} we draw risk neutral PDFs for $r=0.03$ 
and remaining parameters as in figures~\ref{fig:piecewise_panelA} and~\ref{fig:piecewise_panelB}. The effect of the additional terms is evident 
from their comparison. In particular, it is remarkable the increase of the skewness induced by the risk neutral correction from linear plots 
in Panel~(a). Turning our attention to the pricing of Asian options, Figure~\ref{fig:2DPlot_Asian} plots the joint density $p^{\mathbb{Q}}_{UZ}(u, z)$,
and associated marginals for $t_0=0$, $\tau=1$, $\sigma^2=1$, and $\epsilon=2$. Parameters of the fast convolution are 
$z_\mathrm{min}=-10.24$, $m_Z=2^{10}$, $u_\mathrm{min}=-2.56$, $m_U=2^{11}$, and $\Delta\tau=10^{-3}$. 
In figure~\ref{fig:RN-Asian_log} we compare the marginal PDF of $Z_\tau$, and analogously to paragraph \ref{subsec:nr-mnoise} the agreement between FCA and MC is striking
in the central region.

\begin{figure}
	\begin{center}
		\subfloat[]{\resizebox*{0.5\textwidth}{!}{\includegraphics{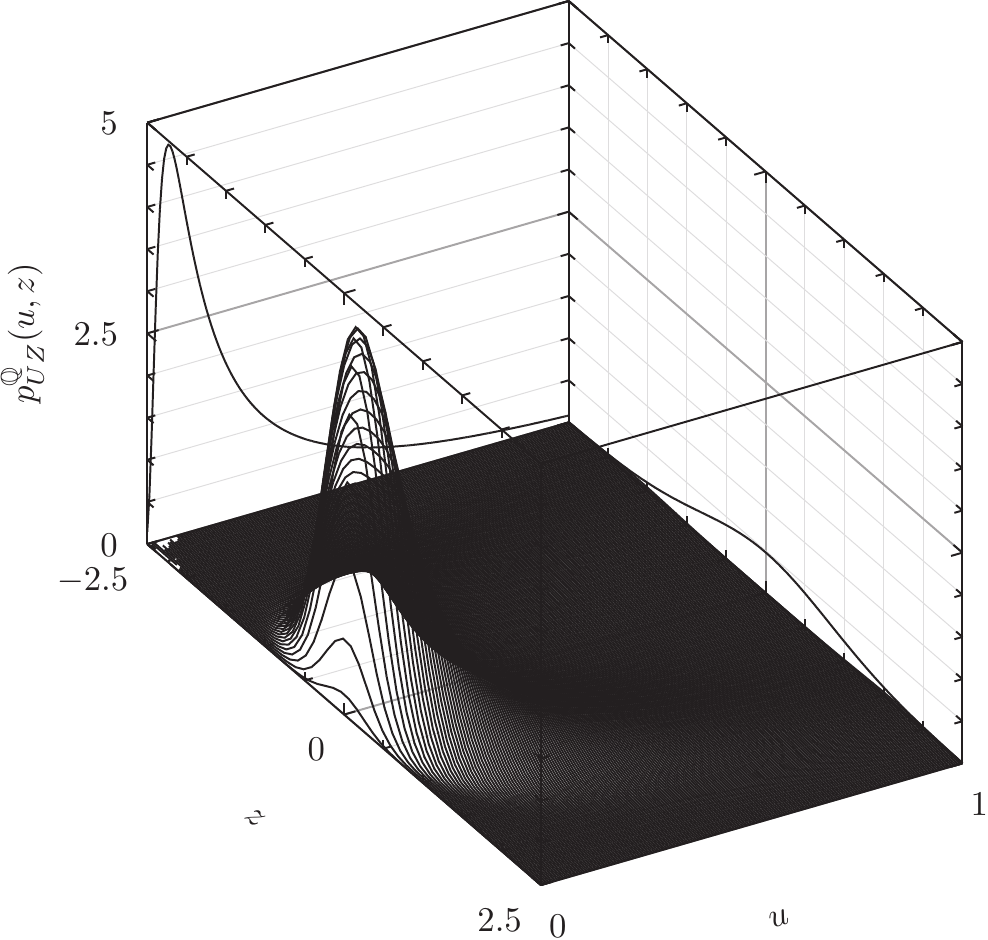}}\label{fig:2DPlot}}%
		\subfloat[]{\resizebox*{0.5\textwidth}{!}{\includegraphics{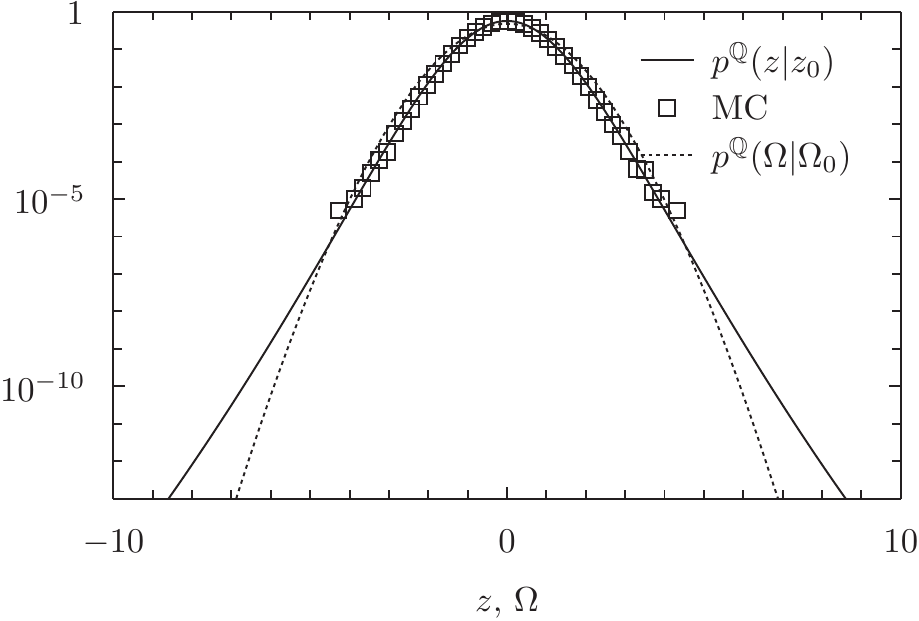}}\label{fig:ZOmega_marginal}}%
		\label{fig:RN_VN}
	\end{center}
	\caption{Vellekoop-Nieuwenhuis-Borland model: 
	Bivariate risk neutral PDF of $Z$ and $U$, and their corresponding marginals; $\alpha=0.1$, $\Omega_0=0$, $t_0=0.2$, and $T=0.7$. 
	In Panel (b) plot of the fast convolution PDFs of $Z$ and $\Omega$  and MC histogram of $Z$.}
\end{figure}

\begin{figure}
	\begin{center}
		\subfloat[]{\resizebox*{0.5\textwidth}{!}{\includegraphics{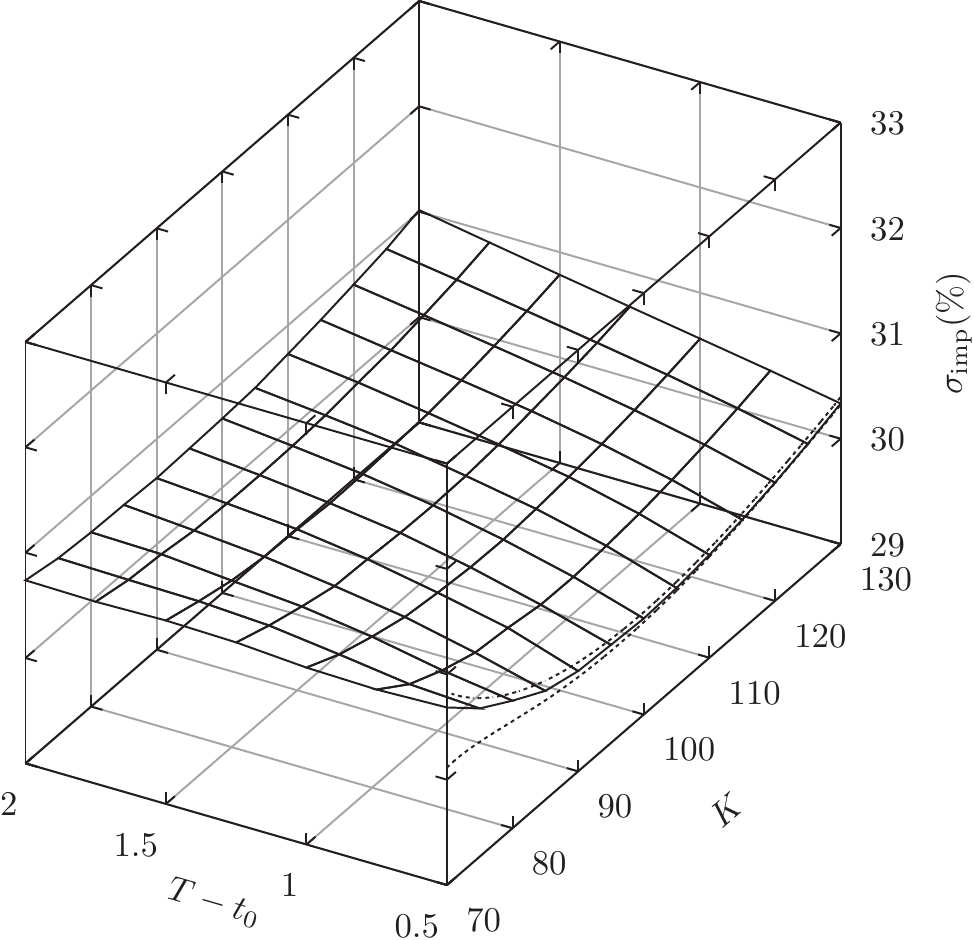}}\label{fig:imp_vola_1_0}}%
		\subfloat[]{\resizebox*{0.5\textwidth}{!}{\includegraphics{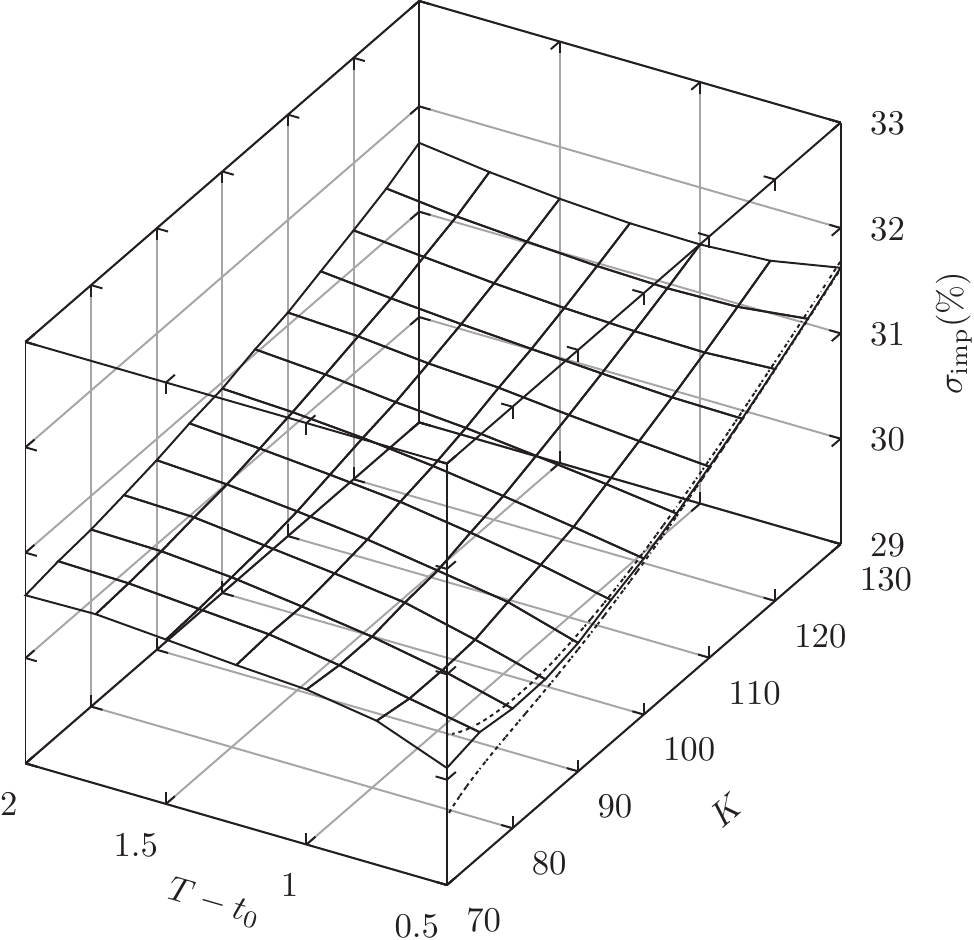}}\label{fig:imp_vola_1_5}}%
		\label{fig:imp_vola_alpha1}
	\end{center}
	\caption{FCA implied volatility surfaces, $\alpha=0.1$, $t_0=0.2$, Panel (a) $\Omega_0=0$, and Panel (b) $\Omega_0=0.5$; 
	dashed lines for $T-t_0=0.5$  correspond to $95\%$ Confidence Level from MC simulation.}
\end{figure}

\begin{figure}
	\begin{center}
		\subfloat[]{\resizebox*{0.5\textwidth}{!}{\includegraphics{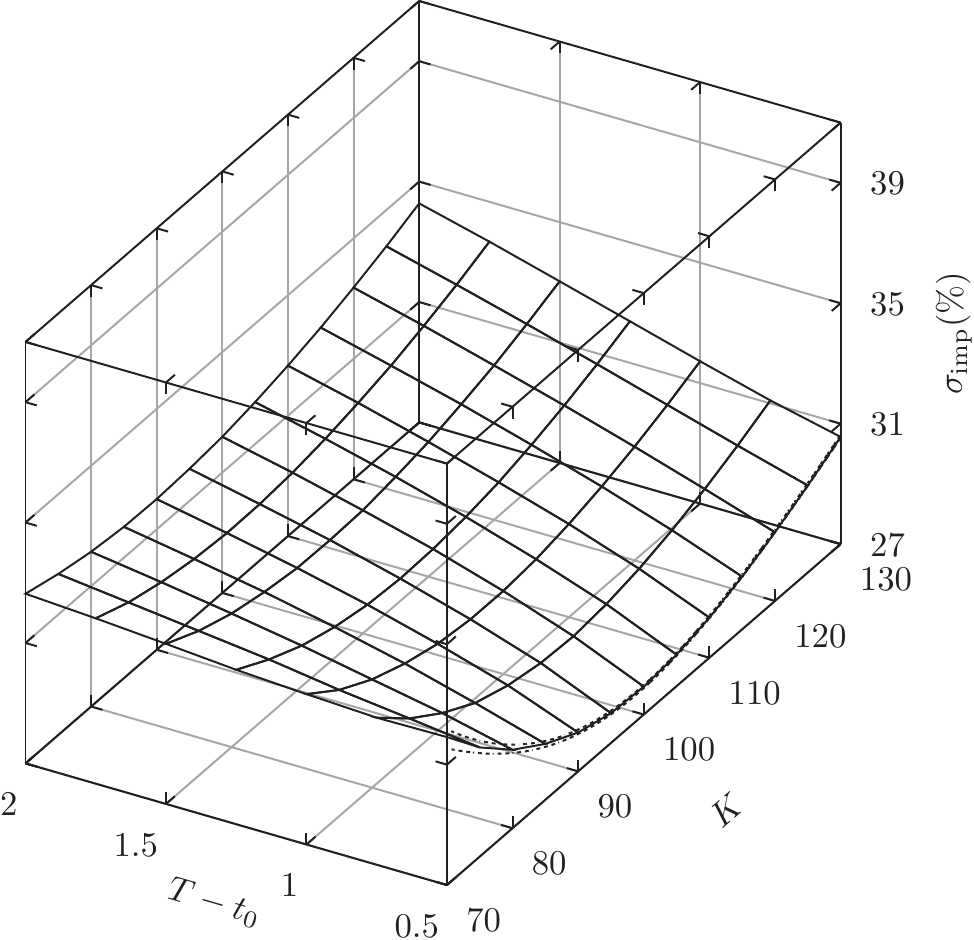}}\label{fig:imp_vola_4_0}}%
		\subfloat[]{\resizebox*{0.5\textwidth}{!}{\includegraphics{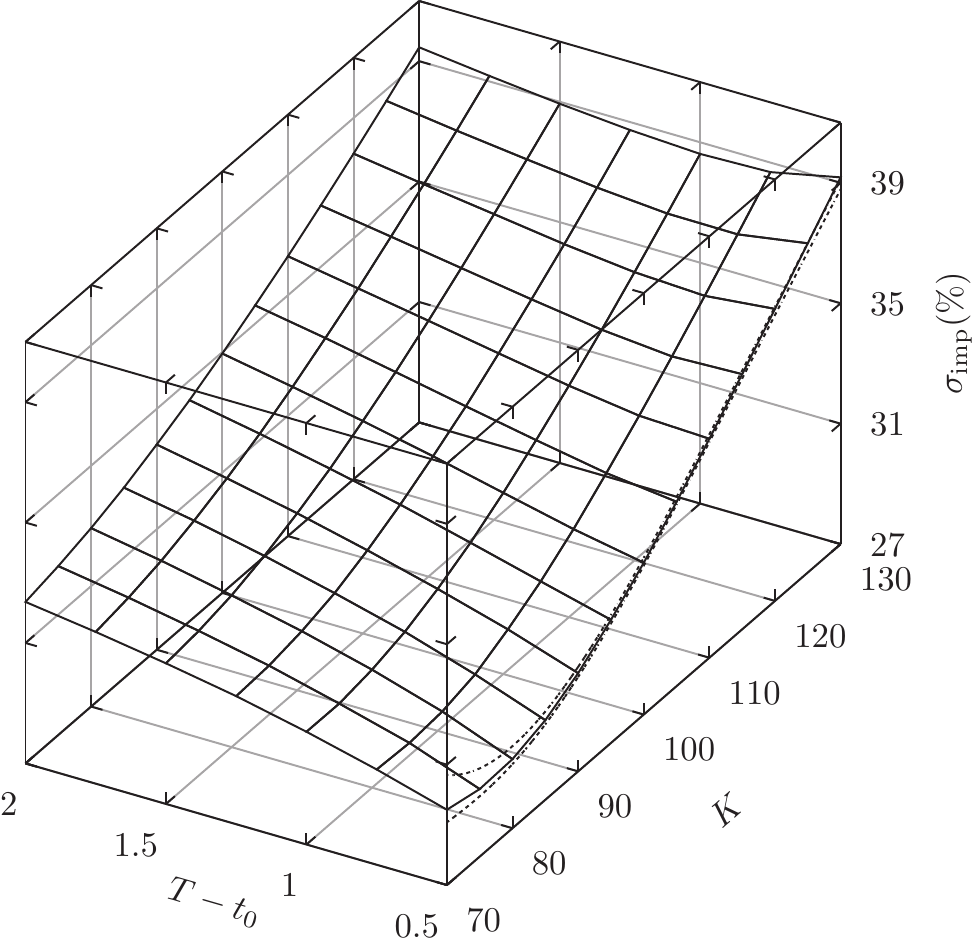}}\label{fig:imp_vola_4_5}}%
		\label{fig:imp_vola_alpha4}
	\end{center}
	\caption{FCA implied volatility surfaces, $\alpha=0.4$, $t_0=0.2$, Panel (a) $\Omega_0=0$, and Panel (b) $\Omega_0=0.5$; 
	dashed lines for $T-t_0=0.5$  correspond to $95\%$ Confidence Level from MC simulation.}
\end{figure}

As far as the pricing under Vellekoop-Nieuwenhuis-Borland model is concerned, we start plotting in figure~\ref{fig:2DPlot} 
the joint bivariate density $p^{\mathbb{Q}}_{U,Z}(u,z)$ for parameter values $z_{{\rm{min}}}=-10.24$, $m_Z=2^{10}$, $u_{{\rm{min}}}=-5.12$, $m_U=2^{11}$, 
$\Delta\tau=10^{-3}$, $\Omega_0=0$, $\alpha=0.1$, $t_0=0.2$, and $T=0.7$. 
We notice that fast convolution algorithm correctly predicts a non negative support for the $U_\tau$ variable, even though the numerical grid spans uniformly the
interval $[u_\mathrm{min},-u_\mathrm{min}]$. In figure~\ref{fig:ZOmega_marginal} we compare the distribution of $Z_\tau$ obtained by means of FCA and MC, 
finding perfect matching, and we also plot $\Omega$ PDF, easily derived given the relationship between the two variables, see equation~(\ref{eq:ZfunctOmega}). 
In light of the agreement between the two numerical procedures, we can use the FCA approach to efficiently price European Call options, 
as explained in paragraph~\ref{subsec:VNB_model}. 
In this respect in figures~\ref{fig:imp_vola_1_0}, \ref{fig:imp_vola_1_5}, \ref{fig:imp_vola_4_0}, and \ref{fig:imp_vola_4_5}
we present our results in terms of implied Black-Scholes volatilities. Our choices of the parameters are $S_{t_0}=100$, $r=0.03$, $\sigma=0.3$, 
$t_0=0.2$, $\Omega_{0}=0,0.5$, $\alpha=0.1,0.4$, $K\in[70,130]$, and $T-t_0\in[0.5,2]$.  MC bands at $95\%$ Confidence Level are plotted as dashed lines 
for the shortest time to maturity, $T-t_0=0.5$ with $N_\mathrm{MC}=5\times 10^7$.  FCA and MC volatility curves are fully consistent. 
As expected surfaces exhibit a volatility smile, more pronounced for small maturities and for $\Omega_{0}$ values deviating from zero. 
As already pointed out by Vellekoop and Niueuwenhuis, a wider variety of volatility surfaces and flexibility of the model can be obtained by playing 
with different values of $\Omega_0$.

\section{Conclusions and perspectives}\label{sec:conclusions}

In this paper we have addressed the problem of investigating performances of the fast convolution algorithm introduced by~\cite{Eydeland}. 
Choosing different specifications of the stochastic process, this has been carried out both with the reconstruction of conditional probability densities 
at different time horizons and with the computation of prices of financial derivatives. FCA is an efficient grid algorithm relying on restating functional integrals 
as sequences of ordinary finite dimensional integrals, and on converting the stochastic process to a unitary diffusion one by means of the Lamperti transform. 
A bright formulation of the problem, then, allows those integrals to be evaluated efficiently by the use of fast Fourier transform techniques.

The stochastic processes we have investigated belongs to two classes of multiplicative noise processes: the family of quadratic diffusion, 
see~\cite{Bormetti_Delpini,Delpini_Bormetti}, and piecewise linear diffusions, see~\cite{McCauley_Gunaratne,Alejandro_etal}.
The analysis performed in this work provides a natural complement to the analytical results obtained in~\cite{Bormetti_Delpini}, where closed form solutions
for the stationary PDF and for the convergent moments at arbitrary time had been obtained. We have detailed a step by step numerical procedure able to provide an accurate
estimate for the probability distribution of the process even far from the stationary regime. Similar results have been found for the piecewise diffusion. 
In this latter case, if the dynamics is enriched with a non trivial drift term, scaling properties are not preserved anymore and every analytical information is lost. 
Being this exactly the situation we faced when switching to the risk neutral setting, FCA proved to be a very efficient and reliable approach to the problem of
option pricing. A detailed empirical analysis for different specifications of the parameter values documents the superiority of the FCA approach 
to standard Monte Carlo simulations. We have also demonstrated the flexibility of the approach when dealing with exotic instruments, and exploited the formal 
analogy between geometric Asian option pricing and Plain Vanilla pricing under Vellekoop-Nieuwenhuis-Borland dynamics. 
Being an interesting hybrid between a geometric Brownian motion and a stochastic volatility model, the latter provides a realistic description of the 
dynamics implied in the option market. FCA is able to numerically reproduce a rich variety of implied volatility surfaces improving the 
standard Monte Carlo approach.

Since, as documented, FCA turns out to be highly successful also in the case of the Vellekoop-Nieuwenhuis-Borland model, a natural perspective is 
to concentrate future research efforts on the extension of FCA to higher dimensional stochastic systems. This is precisely the case of continuous time
stochastic volatility models, see~\cite{Fouque_Papanicolaou_Sircar}. These models provide a flexible framework when modeling volatility, 
and they allow to reproduce several observed statistical regularities. For this reason they are nowadays extensively exploited 
by quantitative sectors of banks and financial institutions. 
Given the ability of the fast convolution to reconstruct densities over tail regions, and of the investigated models to generate leptokurtic and 
scaling distributions, the present approach is naturally suited for application in the context of financial risk management, 
e.g. Value-at-Risk and coherent risk measures computation, see~\cite{Jorion, McNeil_etal, Bormetti_etal_PHYSICAA, Bormetti_etal_JSTAT}.

\section*{Acknowledgements}
The authors acknowledge the support of the Scuola Normale Superiore Grant `Giovani Ricercatori (2011/2012)'.

\end{document}